\documentclass[acmsmall]{acmart}

\renewcommand\footnotetextcopyrightpermission[1]{} 
\setcopyright{none}
\makeatletter
\let\@authorsaddresses\@empty
\makeatother

\usepackage{epsfig}
\usepackage[linesnumbered,ruled,vlined]{algorithm2e}

\usepackage{tikz}
\usetikzlibrary{patterns,matrix,positioning}

\usepackage{bbm,amsmath,amsfonts,latexsym}

\usepackage{listings}
\usepackage{booktabs} 

\usepackage{balance}
\usepackage[linesnumbered,ruled,vlined]{algorithm2e}
\let\chapter\undefined
\usepackage{microtype}
\usepackage{caption}
\usepackage{subcaption}
\usepackage{indentfirst}

\DeclareFontFamily{U}{mathx}{\hyphenchar\font45}
\DeclareFontShape{U}{mathx}{m}{n}{<-> mathx10}{}
\DeclareSymbolFont{mathx}{U}{mathx}{m}{n}
\DeclareMathAccent{\widebar}{0}{mathx}{"73}

\usepackage{graphicx,color}
\usepackage{graphics,url}
\usepackage{multirow}
\usepackage{hyperref}
\usepackage[normalem]{ulem}
\usepackage{tabularx}
\usepackage{array}
\usepackage{hhline}
\definecolor{subtler}{rgb}{1,0,0.1}    
\definecolor{NIST}{rgb}{1,0,0.3}     
\definecolor{NMI}{rgb}{0.8,0.5,0.8}
\definecolor{nicepurple}{rgb}{0.7,0.3,1} 

\def\dv{\color{black}}

\definecolor{NL}{rgb} {0.2,0.7,0.6}  	

\definecolor{negOWD}{rgb}{1,0,0} 

\usepackage{xspace}
\usepackage{textcomp}
\def\ms{$\,\mathrm{ms}$\xspace}         
\def\mus{$\,\mathrm{\hbox{\textmu}s}$\xspace}  

\graphicspath{{.}
{Figures/}
{iHorology/Figures/}
}

\def\wideoneup{177.5mm}
\def\oneup{175mm}
\def\twoup{89mm}
\def\threeup{57mm}
\def\fourup{44mm}

\graphicspath{{.}
{Figures/}
}

\begin{document}

\title{iHorology: Tunnelling through the Barrier to Microsecond-level Internet Time}
\title{iHorology: Sidestepping the Barrier to Microsecond-level Internet Time}
\title{iHorology: Scaling the Barrier to Microsecond-level Internet Time}
\title{iHorology: Dismantling the Barrier to Microsecond-level Internet Time}
\title{iHorology: \  Eroding the Barrier to Microsecond-level Internet Time}
\title{iHorology: \  Lowering the Barrier to Microsecond-level Internet Time}

\author{Sathiya Kumaran Mani}
\affiliation{%
 \institution{Microsoft Research}
 \country{USA}}
\email{sathiya.mani@microsoft.com}
\author{Yi Cao}
\affiliation{%
 \institution{University of Wollongong}
 \country{Australia}}
\email{ycao@uow.edu.au}
\author{Paul Barford}
\affiliation{%
 \institution{University of Wisconsin-Madison}
 \country{USA}}
\email{pb@cs.wisc.edu}
\author{Darryl Veitch}
\affiliation{%
 \institution{University of Technology Sydney}
 \country{Australia}}
\email{Darryl.Veitch@uts.edu.au}





\maketitle
\thispagestyle{empty}
\fancyfoot{}
\subsection*{Abstract} \label{sec:abstract}
 \noindent High precision, synchronized clocks are essential to a growing number of Internet applications.  Standard protocols and their associated server infrastructure have been shown to typically enable client clocks to synchronize on the order of tens of milliseconds~\cite{Hong11}.  We address one of the key challenges to high precision Internet timekeeping - the intrinsic contribution to clock error of {\em path asymmetry} between client and time server, a fundamental barrier to microsecond level accuracy.  We first exploit results of a measurement study to quantify asymmetry and its effect on timing.  We then describe three approaches to addressing the path asymmetry problem: LBBE, SBBE and K-SBBE, each based on timestamp exchange with multiple servers, with the goal of tightening bounds on asymmetry for each client. We explore their capabilities and limitations through simulation {\dv and argument}. We show that substantial improvements are possible, and discuss whether, and how, the goal of microsecond accuracy might be attained.

\vspace{-0.25cm}
\section{Introduction}
Accurate, high precision time synchronization between Internet-connected hosts is essential to a growing number of applications.  Examples include on-line multi-player gaming where event synchronization is critical to quality of experience; financial applications where timing on execution of transactions can have huge implications; network measurement where one-way path measurements of issues like congestion require synchronized hosts; and in distributed databases to enforce consistency.  While Internet time keeping---or iHorology--- has been widely studied, there remain significant challenges in fine-grained clock synchronization at scale.

The standard framework for synchronizing clocks between distributed hosts has three components.  The first is a set of servers with high quality, high precision clocks that act as the authoritative source of time.  The second is the protocol mechanism that specifies the information exchange between the remote server(s) and client(s), and the third is a client synchronization algorithm that adjusts the local clock based on that information, with the objective of achieving tight synchronization between client and server.
 Standard protocols such as the Network Time Protocol (NTP) and its implementations \cite{millsntp,sync_ToN}, that realize this framework have been shown to enable synchronization on the order of a few milliseconds.  


In this paper, we address the problem of client clock \emph{offset} in the Internet, by which we mean the systematic error that a client clock exhibits.    There are two key challenges, {\em (i)} the fact that clients and reference servers communicate over Internet paths that are dynamic in terms of packet latencies and routes that packets traverse, and {\em (ii)} asymmetry in the minimum latencies in the two directions along the path. The first of these is addressed with robust synchronization algorithms, resulting in some residual, hopefully small, average offset error. The second is inherent and cannot be tackled by any algorithm working alone.  In practice these combine to produce an overall offset error, but we posit that the most important of these, which we focus on, is that {\dv due to} path asymmetry.  The broad goal of our work is to enable reliable client clock synchronization on the order of microseconds.  Without tackling the path asymmetry issue, this goal \textbf{cannot} be {\dv approached, much less} achieved.

To justify our claim of the importance of path asymmetry, we provide a basic model for network clock synchronization that shows that even under idealized conditions, asymmetry translates directly to offset estimation errors.  We then use a unique reference dataset \cite{sync_TimeServer_Dataset2016-17} with high-precision One Way Delay (OWD) measurements to deepen insight into the issue, by identifying and characterizing path asymmetry in the Internet.  We find that path asymmetries are common, significant and variable, and that their expected impact on offset estimation is typically greater than 1\ms and can be as much as 100's of \ms.  Our empirical analysis serves as the foundation for an asymmetry model that can be used in simulations, and, in a unique approach, to gain insights into asymmetry impacts.  In particular we explain how \emph{asymmetry jitter}, which is caused by a client changing its reference server, can degrade clock offsets to a significant degree, an effect that
increases with the number of servers used.

Next, we describe a general approach for reducing the impact of path asymmetry, by 
utilizing information from multiple servers simultaneously to establish {\em tighter bounds} on asymmetry.  We utilize both a model and measurement data to show that estimation bounds tighten as the number of contacted servers grows, however the extent of tightening is modest, and depends strongly on which servers are included.

We address what would appear to be a fundamental limitation to further reduction of offset errors by utilizing independent, a priori information about the network.  Our first approach, SBBE, is deterministic. It takes advantage of a speed-of-light (SoL) delay approximation between a client and multiple servers to establish strict lower bounds on OWDs.  The second approach, LBBE, inspired by~\cite{Wong07}, is statistical. It combines SoL delay with measured OWDs between landmark servers to learn delay bound estimates in the network, leading to tighter, but less certain bounds for clients.  
Our third approach, K-SBBE, improves SBBE by incorporating weak routing information.

We use {\dv simulation to evaluate our methods, based on models reflecting fundamental principles of asymmetry and its impact on timing.}
Simulation is required because of the near impossibility of controlling a non-trivial population of geographically distributed clocks (let alone the 1050 we use), each with the reference timing required for validation.
{\dv The nature of our goal, the evaluation of fundamental limits and impacts, and pathways to their mitigation, is well suited to the simulation setting, as detailed network and traffic models are not needed to achieve robust insights.}
Our results show that each method {\dv has the potential to} achieve significant improvements in offset estimates compared to a protocol that does not consider path asymmetry.  We show that this can be done with only a modest number of measurements, and by clients contacting only a modest number of servers, and show that the results are robust in the face of congestion induced variability.


Throughout, we describe why and to what extent our methods fall short of the goal of microsecond level timing. We conclude by assessing how close to the goal we have come, and how the remaining gap might be bridged.
This work does not raise any ethical issues.

\vspace{-0.2cm}
\section{Internet Timing Background}
We focus the discussion on the Network Time Protocol (NTP),  since it is by far the most widely used timing protocol in the Internet.  The original version of NTP~\cite{rfc958} enabled clock synchronization on the order of 10's to 100's of milliseconds.  The current standard for NTP is described in RFC 5905~\cite{rfc5905}.  

NTP uses a hierarchy of clocks to provide timing information to end-host clients.  At the top are the stratum-0 references: stable, high-precision clocks based on GPS-based receivers or atomic sources. These are not networked servers, rather they act as reference sources of Universal Coordinated Time (UTC) to  stratum-1 servers through direct, local connections.  For $s=1,2,\ldots 15$, stratum--$(s\!+\!1)$  servers synchronize over the network to stratum-$s$ servers.
Clients typically connect to stratum-2 or 3 servers, and often connect to multiple servers in an attempt to improve synchronization, or for redundancy.  
The independent, but widely used NTP pool server infrastructure~\cite{ntppoolproject} is designed so that a convenient DNS request for \texttt{pool.ntp.org}  results in multiple server IP addresses that are geographically `close' to the client.

NTP client software, \textit{ntpd}, is included in most operating system distributions.
It maintains a periodic request-response packet exchange `polling' reference servers, each resulting in four timestamps, recording when {\em (i)} the client request is sent, {\em (ii)} is received by the server, {\em (iii)} is sent back from the server, and {\em (iv)} is received at the client.  These values are used by ntpd heuristics to attempt to drive clock error to zero.  Polling frequency is determined by the measured round-trip delay, jitter and local oscillator frequency.

Client clock errors can be introduced by a variety of factors that fall into four categories:  quality of client oscillators, variability of network path latencies, dynamics of the clock synchronization algorithm, and underlying path asymmetry, which is the focus of our study.

\section{Related Work}
\label{sec:related_work} Internet Clock synchronization has been studied for many years  (as a small sample, see \cite{Marzullo83,mills81dcnet,mills85alg}), yet despite its importance as a {\dv major} source of error, many works do not even mention {\dv path} asymmetry, few address it, and very few address it in any meaningful way with reliable measurements or analysis.  
None that we are aware of combine theoretical understanding with practical approaches to tackle asymmetry issues in Internet timing practice.

Internet measurement literature: \ 
Internet routing asymmetry at the IP level has  been studied, for example \cite{Fraleigh2003,Zeitoun2004,He05,Schwartz10,John2010,Wassermann2016}. One of the first was Paxson~\cite{Vern97}, who used IP addresses returned by TTL-limited probing to identify asymmetries between a set of deployed measurement nodes.  
Pathak {\em et al.}~\cite{Pathak2008} study Internet path asymmetry using OWD measurements in addition to IP level routing.  However their study is inherently limited by the accuracy of the NTP-based time synchronization on their measurement nodes.
Lin {\em et al.} examined data from 63 (resp.~361) 
stratum-1 servers with a polling period of 90s (resp.~90min.)  over 24hrs (resp.~115hrs)~\cite{Hong11}.  They found that NTP enables synchronization with a median error of 2--5\ms but that error can often be much greater.  They also showed that inaccuracies were due primarily to server inaccuracy, path asymmetries and queuing delays, however this was not based on any methodology enabling congestion, routing and server effects to be disambiguated.
In contrast, our study uses a much larger and more authoritative dataset of synchronized nodes \cite{sync_TimeServer_Dataset2016-17}, at 1 second granularity, to study asymmetry in detail. This dataset has previously been used to study the quality of stratum-1 servers \cite{sync_Leap2016,sync_Best50}. 
We are the first to exploit it to survey path asymmetry. More generally, we are the first to measure asymmetry while accounting in a principled way for the strongly distorting effect of routing changes, and the first to propose a model for empirical Internet asymmetries.

Synchronization analysis literature:  \ 
Marzullo and Owicki  \cite{Marzullo83} were the first to suggest that that intersecting clock error intervals derived from neighboring clock comparisons could be used to improve local estimates. This was refined by 
Freris {\em et al.}~\cite{Kumar_TAC2010} who derived fundamental limits to joint synchronization of a network of affine clocks, and related results were obtained by Huygens~\cite{huygens} in the context of Data Center synchronization. 
In contrast, our work develops a model that is fully consistent with the general work of \cite{Kumar_TAC2010}, but is focused on the individual client viewpoint, using assumptions and solution approaches that are Internet-aware and capable of addressing the practical problem of asymmetry mitigation for Internet timekeeping.
They could be incorporated into existing  synchronization algorithms such as Huygens, NTP, RADClock~\cite{sync_ToN} and even potentially 
SPoT~\cite{SpotIOT18}.

\section{Asymmetry and Timing}
\label{sec:asym}
\newcommand{\be}{\begin{equation}}
\newcommand{\ee}{\end{equation}}
\newcommand{\ben}{\begin{equation*}}
\newcommand{\een}{\end{equation*}}
\newcommand{\ba}{\begin{eqnarray}}
\newcommand{\ea}{\end{eqnarray}}
\newcommand\Var {{\rm Var}}
\newcommand{\Z}{Z\!\!\!Z}
\def\integer {\cal Z}
\def\real {{\mathcal R}}
\def\bl{\Bigl(\,}
\def\br{\,\Bigr)}

\newcommand{\HRule}{\noindent\rule[2mm]{\linewidth}{0.4mm}\vspace{-2mm}}
\newcommand{\tinyskip}{\vspace{0.5mm}}
\newcommand{\miniskip}{\vspace{1mm}}
\newcommand{\ntinyskip}{\vspace{-0.5ex}}
\newcommand{\nminiskip}{\vspace{-1ex}}
\newcommand{\nsmallskip}{\vspace{-2ex}}
\newcommand{\nmedskip}{\vspace{-3ex}}
\newcommand{\nbigskip}{\vspace{-5ex}}

\def\wideoneup{177.5mm}
\def\oneup{175mm}
\def\twoup{89mm}
\def\threeup{57mm}
\def\fourup{44mm}


\newcommand{\dbb}{\underbar{d}^\downarrow_L}
\newcommand{\dfb}{\dfm_L}

\newcommand{\sdm}{\underbar{d}^\rightarrow\!}
\newcommand{\dfm}{\underbar{d}^\uparrow}
\newcommand{\dbm}{\underbar{d}^\downarrow}
\newcommand{\Rm}{\underbar{r}}
\newcommand{\Am}{\underbar{a}}
\newcommand{\ra}{T}

\newcommand{\Dfi}{D^\uparrow_{i}}
\newcommand{\Dbi}{D^\downarrow_{i}}
\newcommand{\Sdi}{D_i^\rightarrow}

\newcommand{\erm}{\hat\Rm}
\newcommand{\eam}{\hat\Am}
\newcommand{\dfh}{\widehat{\underbar{d}}^\uparrow}
\newcommand{\era}{\widehat{\ra}}

\newcommand{\df}{d^\uparrow}
\newcommand{\db}{d^\downarrow}
\newcommand{\sd}{d^\rightarrow}
\def\al{a_l}
\def\ar{a_r}
\def\ac{a_c}
\def\cl{c_l}
\def\crr{c_r}
\def\tb{\widebar t}
\newcommand{\qf}{q^\uparrow}
\newcommand{\qb}{q^\downarrow}

\newcommand{\Tr}{\mathbf{\ra}}

\def\simpleroutemodel                   {\includegraphics[width=80mm]{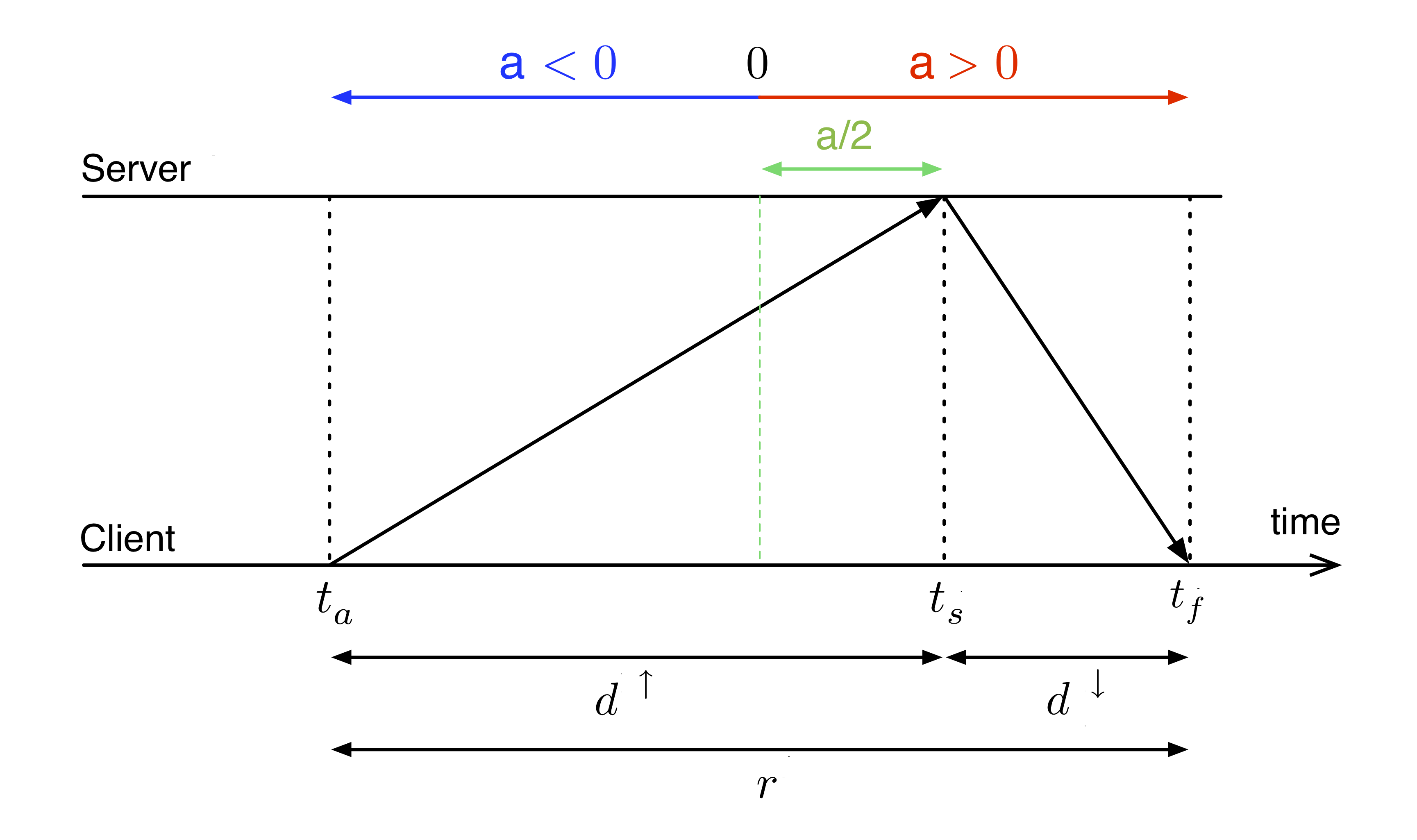}}
\def\RAscatter                     {\includegraphics[width=145mm]{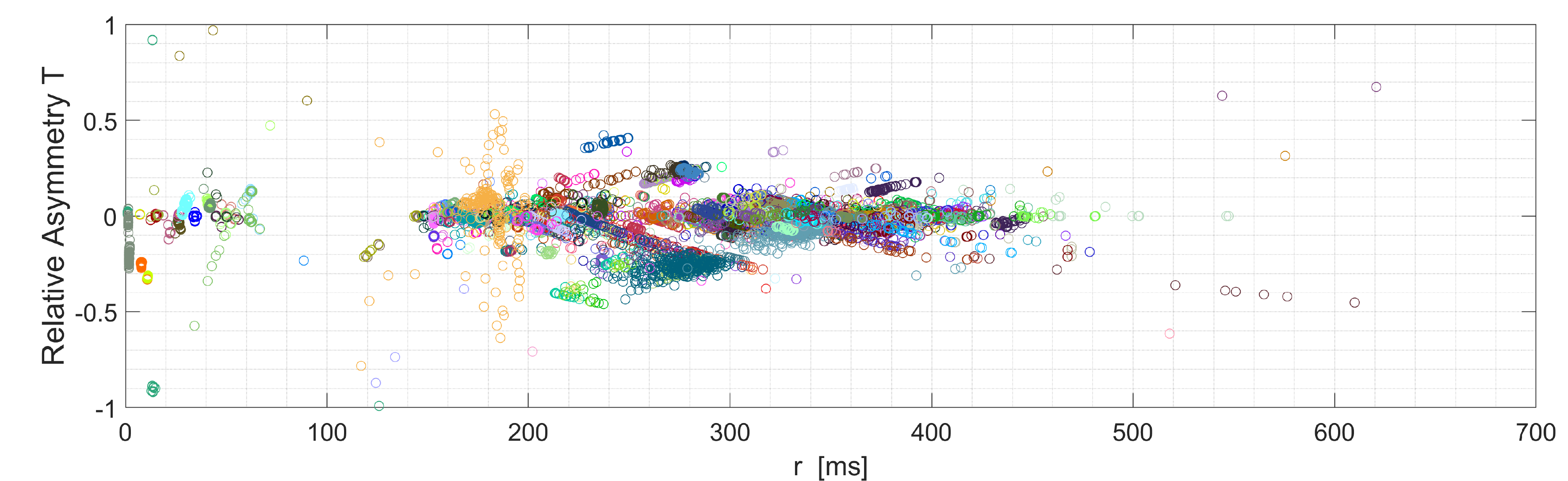}} 
\def\ECDF			                    {\includegraphics[width=88mm]{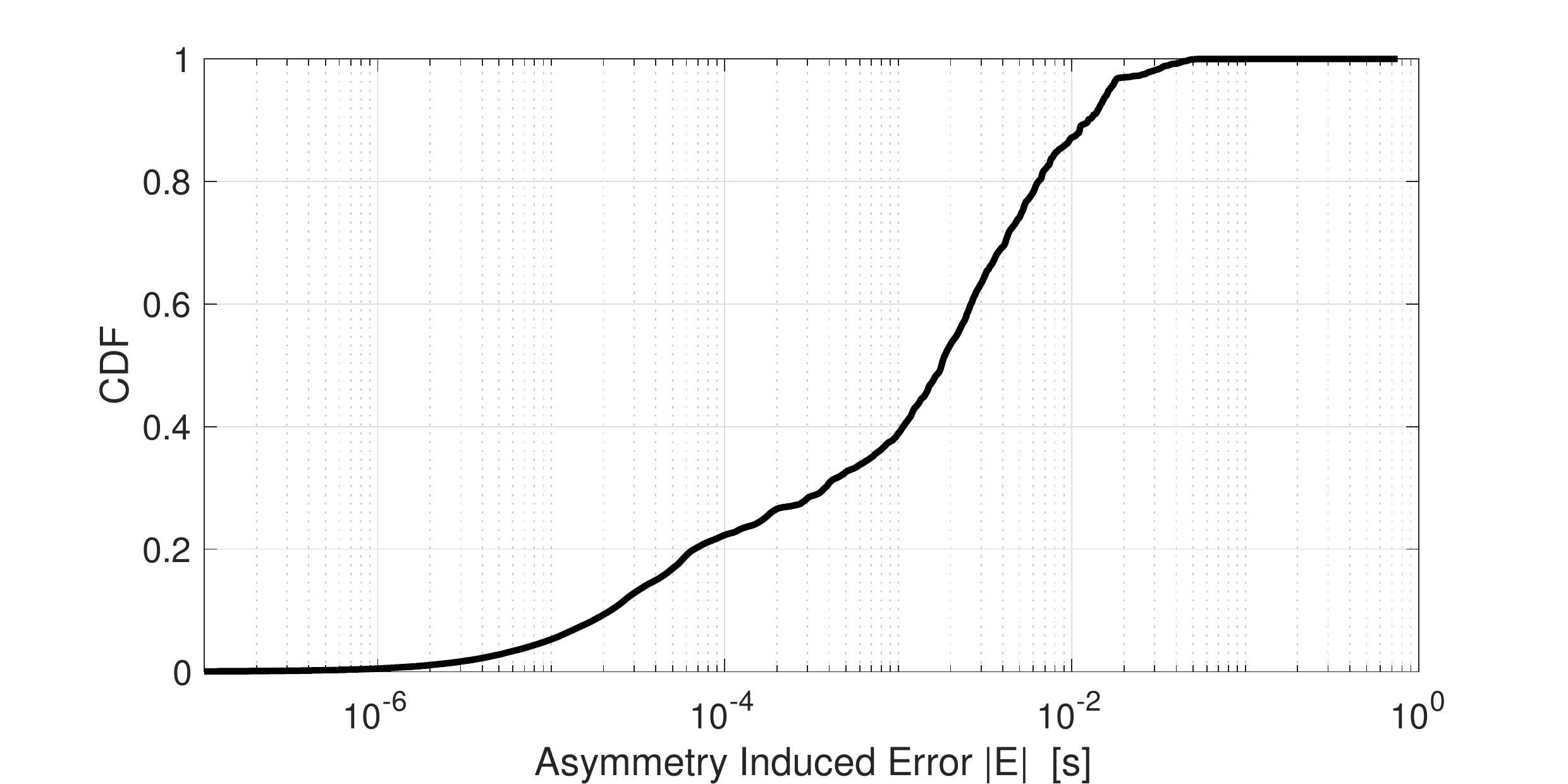}}
\def\RACDFandmodel                   {\includegraphics[width=93mm]{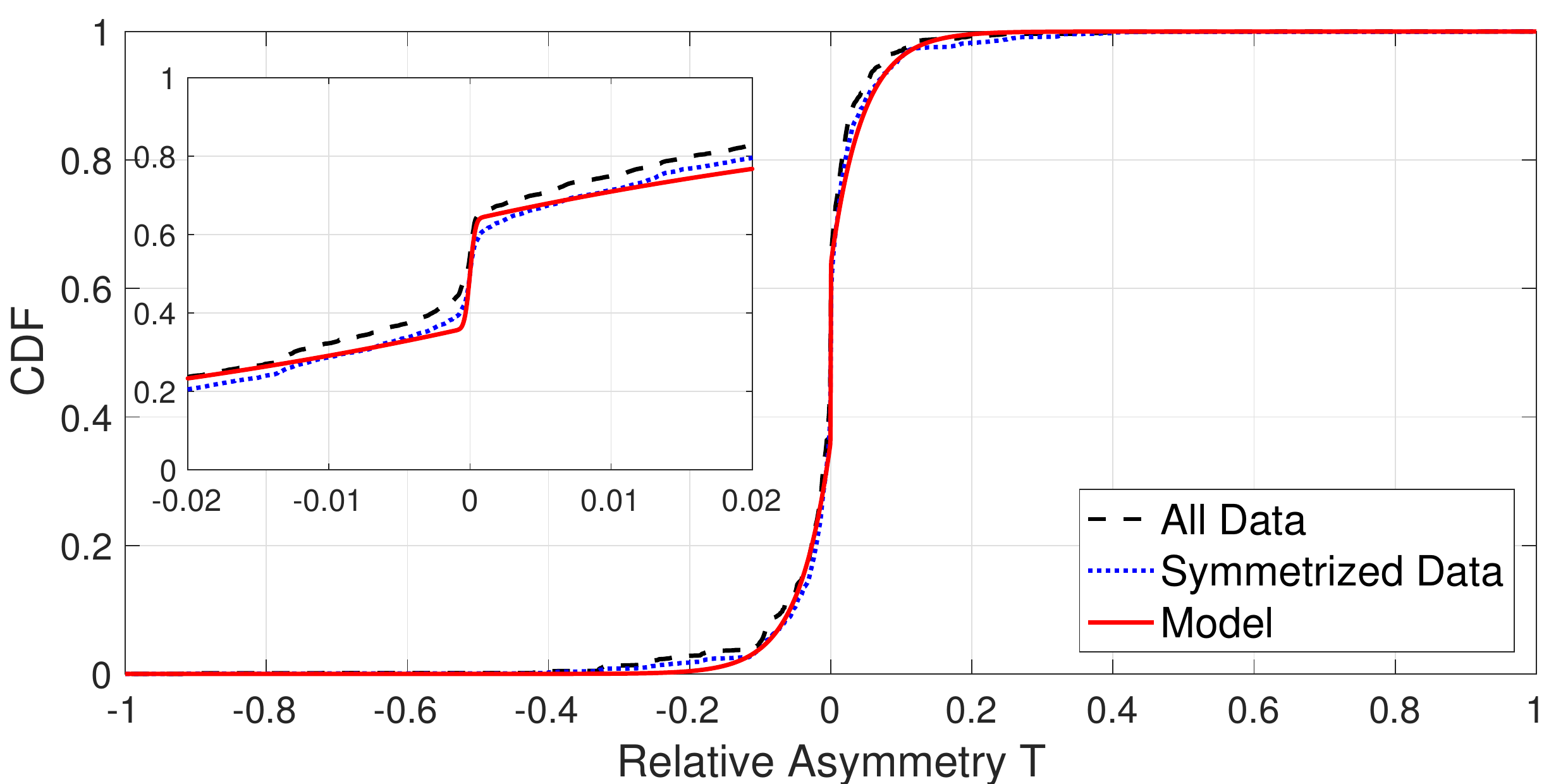}}   
\def\jitterError			            {\includegraphics[width=103mm]{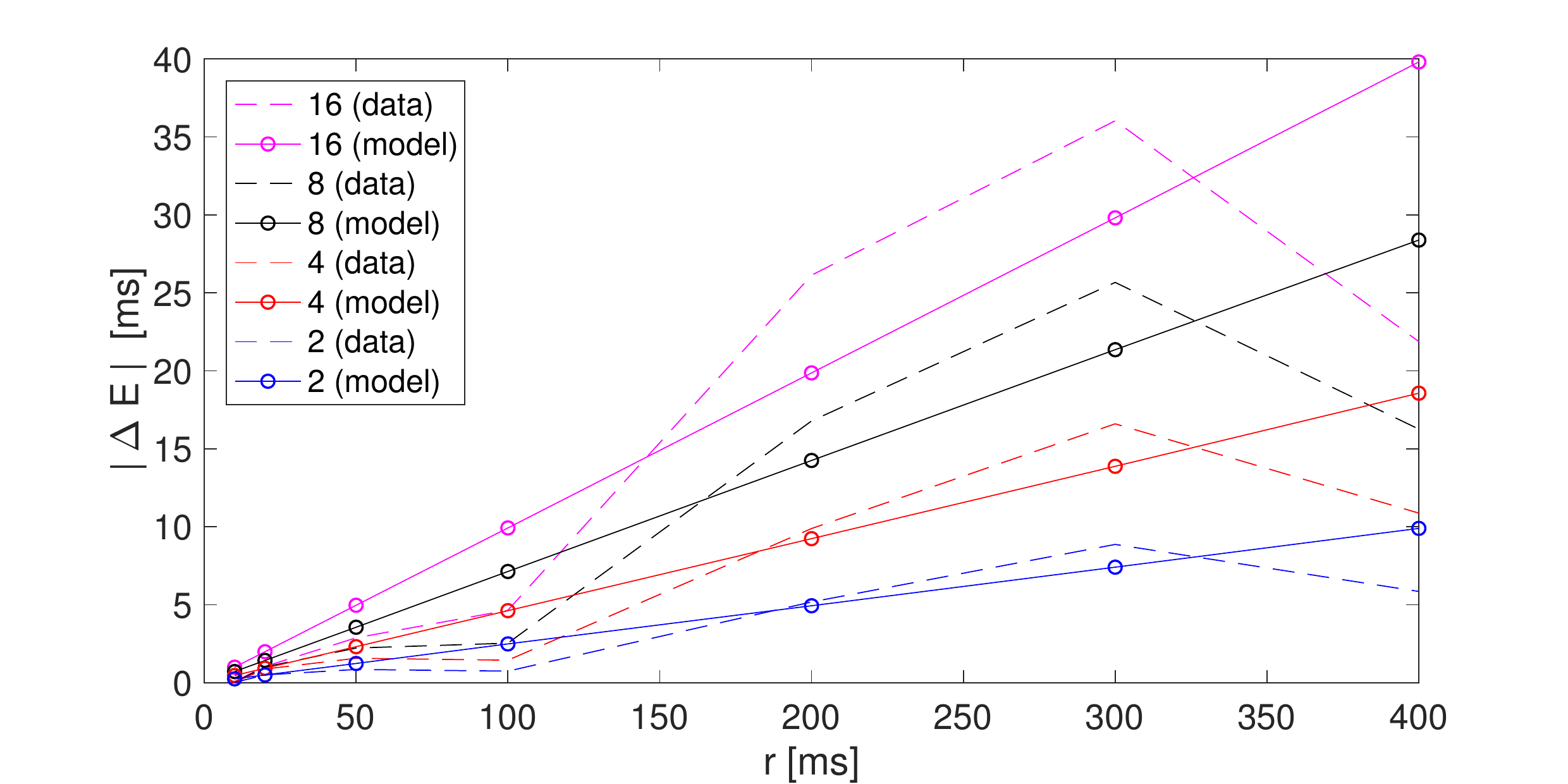}}   
\def\boundtightening			    {\includegraphics[width=90mm]{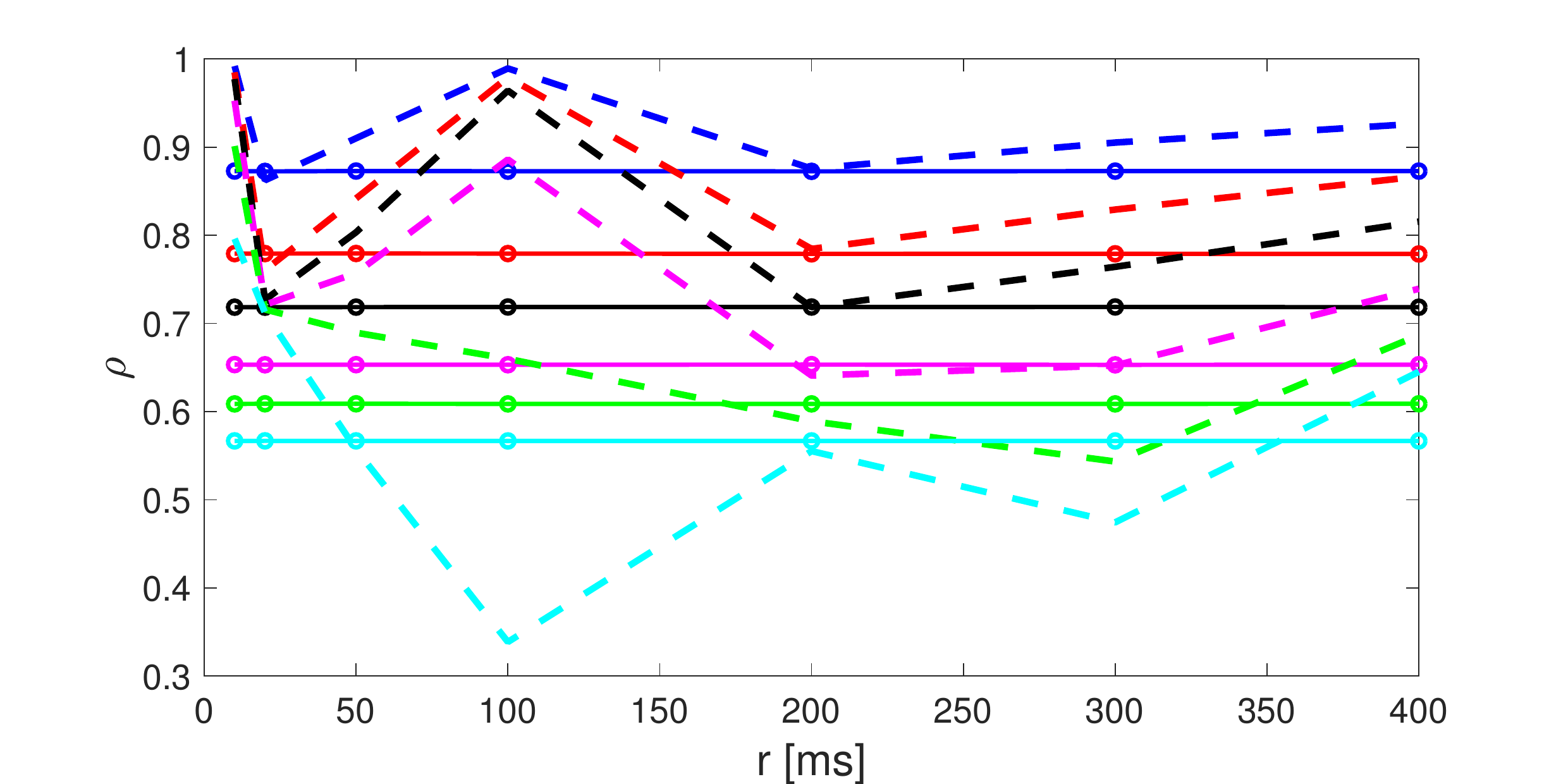}}


In this section we describe the underlying model of network clock synchronization, explain why path asymmetry is so important, and why is it a difficult problem.  We exploit a unique dataset to make reliable measurements of asymmetry found in the Internet, and then exploit these measurements in three ways:  {\em (i)} to propose a model that describes observed asymmetry and informs our simulations, {\em (ii)} to illustrate the reality and dangers of \textit{asymmetry jitter}, and {\em (iii)} to demonstrate the principles of {\dv how  asymmetry diversity can be exploited as} a resource to reduce impacts on timekeeping, and to evaluate its potential.

\subsection{Fundamentals}
\label{ssec:fund}

\begin{figure}[b]
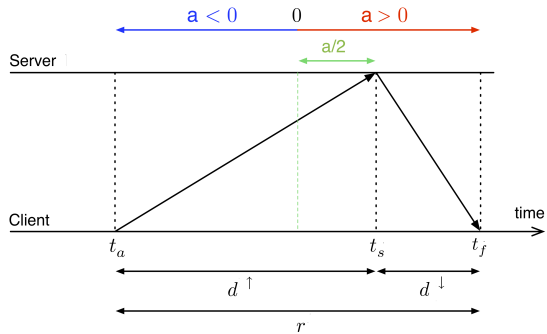

     \vspace{-1mm}
   \begin{center}
       \simpleroutemodel
    \end{center}
    \caption{The route model with positive asymmetry. \dv In practice the server event splits into arrival and departure.}
    \label{fig:routemodel}
\end{figure}

Consider a client clock $C(t)$, with error $E(t) = C(t)-t$, and a server clock $S(t)$, where $t$ is the true time.
The fundamental idea in remote synchronization using a round-trip c$\rightarrow$s$\rightarrow$c
timestamp exchange, illustrated in Figure~\ref{fig:routemodel}, is to adjust $C$ so that the server timestamp lies inbetween those of the client, that is:  
$C(t_a)<S(t_s)<C(t_f)$ (for simplicity, we neglect the delay at the server).  
The question is, where exactly should it be put?
The correct answer, which would reduce $E(t)$ to zero, is such that after correction
$S(t_s)-C(t_a) = \df$, and $C(t_f)-S(t_s)= \db$, where $\df$ and $\db$ are the forward (c$\rightarrow$s) and backward 
(s$\rightarrow$c) delays respectively.  However the client cannot do this as it cannot measure these delays.
It can only observe the 
{\dv
\textit{measured delays}  
$D_f=\df - E(t)$, and $D_b=\db+E(t)$.
A measured forward delay of $D_f$ is fully consistent with a family, parameterized by $\tau$, of possible clock errors $E(t)-\tau$ and matching measured delays $\df-\tau$, 
all equally consistent from $c$'s viewpoint since 
$D_f = \df - E(t) =(\df-\tau) - (E(t)-\tau)$
(the backward case is similar).}


To reduce the problem to its essence we make three simplifying assumptions:  zero path variability, a perfect server clock, that is $S(t)=t$, and a client clock which is perfect up to a constant offset. This corresponds to an ideal situation of no routing changes, and a synchronization algorithm which has perfectly filtered path variability.
The forward and backward delays then reduce to their minimum values:   $\dfm$, $\dbm$ respectively, as does the RTT: $\Rm=\dfm+\dbm$. 
The underlying constant path \textit{asymmetry} is just $\Am=\dfm-\dbm$, and takes values in $[-\Rm,\Rm]$.  

The client clock model and its error
now reduce to 
\be
   \label{eq:C}
   C_a(t)= t + (\Am-a)/2,   \quad  E_a = C_a(t)-t = (\Am-a)/2,
\ee
where $a$ is a clock configuration parameter specifying where,
relative to the midpoint of $[C(t_a),C(t_f)]$,  the server timestamp is deemed to lie.
To ensure (apparent) causality is respected, i.e.~$C(t_a)\le S(t_s)\le C(t_f)$, we must select $a\in[-\Rm,\Rm]$.
If an estimate of $\Am$ were available, then $a$ could be set to it, but typically 
nothing specific is known, and $a=0$ is used, inducing
an error $E_0=\Am/2$ that is unknown to the client. 

The importance of asymmetry to clock accuracy is now clear.  Even for a perfect client algorithm, with the default $a=0$ the clock error is $\Am/2$, and since $\Am\in[-\Rm,\Rm]$ and $\Rm$ can be 10's or even 100's of milliseconds, asymmetry has the potential to hold clock error forcibly beyond the reach of the goal of 1\mus, by orders of magnitude.
For the imperfect algorithm the situation is of course worse. The 
clock now includes not only the inherent error due to asymmetry, but an additional systematic offset error $\theta$, idealizing algorithmic limitations.  Although the client cannot distinguish these, the oracle can write the total error as
\be
	E_a = \theta + (\Am-a)/2 \ .
	\label{eq:offsetmodel}
\ee
In real networks path delays are not constants, but the underlying values 
$\dfm$, $\dbm$, $\Rm$  and $\Am$  continue to be central, with the only new element compared to the model above being that $\Rm= \dfm+\dbm+\sdm$, where $\sdm$ is the minimum of
the delays $\{\sd\}$ within the server separating its incoming $s_i$ and outgoing $s_o$ timestamps (this is usually a few 10's of \mus).

\subsection{Asymmetry in the Wild}
\label{ssec:wild}

To evaluate the impact of asymmetry induced error, we must know the extent of client$\leftrightarrow$server path asymmetries in practice.   This is difficult to determine universally, as it depends on their round-trip paths.  It is very difficult practically, as it requires absolutely synchronized clocks at each end to measure. 
To gain insight into this question, we make use of the timeserver dataset available at \cite{sync_TimeServer_Dataset2016-17}, and described in \cite{sync_Leap2016}.
The dataset contains round-trip timestamp data collected over a period of two months between a single client (load balanced over two hosts) in a Sydney based testbed with reference timestamping (GPS+atomic+hardware capture card), to some 479 NTP servers spread globally, of which 462 are Stratum-1.  The NTP packet exchange between each (client,\,server) pair runs independently in parallel, generating $\{C(t_a),S(t_{s_i}),S(t_{s_o}),C(t_f)\}$ timestamp 4-tuples once per second (in most cases). 

To exploit this rare opportunity to measure asymmetry, the challenge posed by routing must first be overcome. Each path change results in a different value of $(\dfm,\dbm)$, and hence of $(\Rm,\Am)$. 
Over two months 10's to 100's of such changes occur in a typical server, with $\Delta{\Rm}$ varying from a few \ms, to 10's of \ms, to even over 100\ms.  To measure $(\Rm,\Am)$ reliably, we must identify \textit{Clear Zones} (CZ) free of such changes.

\miniskip\noindent\textbf{Measurement Methodology} \quad
Our approach is built around \textit{LSD}, a Level Shift Detector which detects changes in level 
in a timeseries obeying a model $x(i) = b(i) + q(i)$, where $b(i)$ is a piecewise constant \textit{b}aseline, and $q(i)$ is a positive congestion term.
Each of $\df$, $\db$, and the RTT $r$ can be represented by this model under routing changes.  
LSD implements a sliding window search to locate shifts above an input threshold $\lambda$.
The window size is determined adaptively through a false-positive probability parameter $\alpha$, using an independent, identically distributed model for $x(i)$ between shifts, calibrated using the observed threshold probability $p(\lambda)=\Pr\{x(i)<b(i)+\lambda\}$. 

Based on LSD, we define a Clear Zone Detector (CZD) over a block that, for  $\df$ and $\db$ separately, applies LSD (at fixed $\alpha$) for three different values of $\lambda$ corresponding to $p(\lambda)=\{0.02,0.05,0.2\}$.  
The block is marked as a CZ candidate only if no shifts are found across $\df$ and $\db$, for each $p$. Multiple $p(\lambda)$ values ensure the detector is sensitive despite the conservative ({\dv very} small $\alpha$) false positive control, as it generates a diversity of window sizes, increasing robustness w.r.t.~distributional variation and congestion timescales.

Our approach is as follows.  
In a first phase the data is split into 200 blocks, to each of which CZD is applied with $\alpha=10^{-6}$, and successful blocks labelled as CZ candidates. 
Although this harvests very wide CZs efficiently, unsuccessful blocks may nonetheless contain many more.  
To harvest these, in a second phase we use LSD (with $(\alpha,\lambda)=(10^{-6}, 10^{-4})$) in a new way: 
using its detections to directly partition each failed block into subblocks.   
Each such subblock is then tested with CZD as before, with successful ones added to the candidate pool.  
In a final phase, data samples where the server's stratum was not 1 are removed. 
The final CZ set consists of those candidates with at least 100 samples remaining.

The method was tested and tuned on 5 servers, generating CZs which {\dv collectively} covered over 50\% of the trace durations on average, and were found to be correctly classified upon manual inspection.
The actual proportion of clear zones is without doubt higher, but the current harvest, conservative by design, is adequate for a useful assessment of asymmetry variation. 

As pointed out {\dv forcefully} in \cite{sync_SHM}, \cite{sync_Leap2016}, even stratum-1 servers may have faulty timestamps.  We believe that our CZ detector, in addition to excluding most routing events and zones of high congestion where estimation would be unreliable, will be effective in screening out most such server errors if present.
All zones intersecting the server errors {\dv exhaustively} catalogued in `tag files'  
at \cite{sync_TimeServer_Dataset2016-17} have also been removed.
Finally, the end-2016 leap second occurred within the data.  Periods of related erroneous behaviour exhibited by some servers, as defined in \cite{sync_SHM}, have been excluded.

For each CZ, we estimate $(\Rm,\Am)$ as
\ntinyskip
\ba
   \erm &=& \min\left\{ \min_{CZ}  \df  + \min_{CZ} \sdm  + \min_{CZ} \db , \ \min_{CZ} r \right\} \\
   \eam   &=& \min_{CZ}  \df  -  \min_{CZ} \db \ .
\ea

\smallskip\noindent\textbf{Results} \quad  We measured a total of 153,737 CZs over the 462 stratum-1 servers, with the number of zones per server ranging from $0$ (some servers have continuous errors and/or a high density of path changes), to $1305$.

\begin{figure}[b]
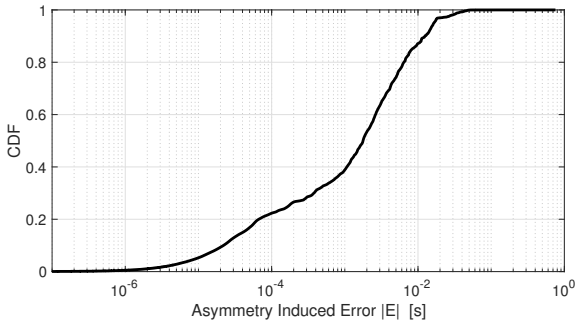

    \vspace{-1mm}
    \begin{center}
       \hspace*{-0.2cm}\ECDF
    \end{center}
    \vspace{2mm}
    \caption{CDF of absolute clock error $|E|=|\Am|/2$ (with $a=0$).}
    \label{fig:ECDF}
\end{figure}

To see the implications for clock error, Figure~\ref{fig:ECDF} gives the CDF of the associated absolute errors $|E|= |\Am|/2$ in log scale. 
Only 0.50\% of the asymmetries imply errors below 1\mus, 
only 4.1\% below 10\mus, and 
only 40.0\% below 1\ms.  

To gain insight into asymmetry independently from $\Rm$, it is natural,
since $\Am\in[-\Rm,\Rm]$, to work with the normalized \textit{relative asymmetry} $\ra= \Am/\Rm\in [-1,1]$, which we estimate via
\be
	 \era =  {\eam}\,{\large/}\,{\erm\ }  \ .
\ee

\begin{figure*}[t]
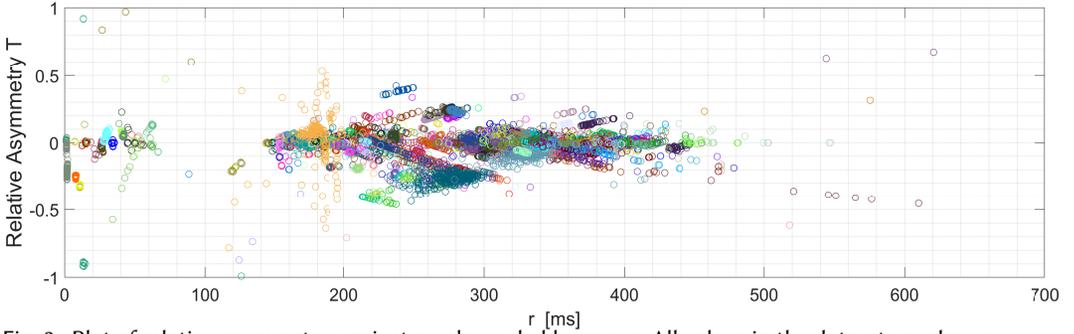

    \hspace{-6mm}   \RAscatter     
    \vspace{1mm}
    \caption{Plot of relative asymmetry against $\Rm$, color coded by server.  {\dv All values in the dataset are shown}.
    }
    \label{fig:scatter}
        \vspace{0mm}
\end{figure*}

A scatterplot of $\era$ versus $\erm$ over all CZs is given in Figure~\ref{fig:scatter}, where CZs from the same server share a common color.
The spatial specificity of the data is immediately apparent:  the gap centred on $\Rm=100$\ms reflects the separation of Oceania from the rest of the world,  and $\Rm=280$\ms essentially delineates the servers in the Americas from those in Europe. 
Apart from this variation in data availability however, the spread of $\ra$ is remarkably independent of $RTT$. 
We observe in particular an overall symmetry about $\ra=0$ regardless of $\Rm$, reflected in a median of just $-2.2\times 10^{-4}$.
This is expected due to the roughly symmetric nature of layer 2 and 3 networks, and to a lesser extent IP routing. More generally however, since the roles of client and server can always be reversed, a symmetric distribution of $\ra$ over $[-1,1]$ is expected in principle.

There is marked variation in the CZs both across servers, and within them.  Within servers a variety of patterns are seen, including {\dv clusters,} and vertical and `linear' structures corresponding to multiple routing changes which affect both, or just one direction to the server.
The scatterplot hides \textbf{many} overlapping points, overemphasizing outliers. To summarize the plot,
the CDF of all $\era$ values, given in Figure~\ref{fig:RACDF} (dashed black curve), reveals a close to symmetric distribution featuring a remarkably sharp peak about $\ra=0$.
The peak is also a consistent feature of subsets based on stratifying with respect to $\Rm$.
{\dv For interest, two zooms of the scatterplot are made available in the Appendix where the intra-server and inter-server structures are visible in greater detail.}

\subsection{A Simple Model for Real Asymmetries}
\label{ssec:model}

We describe a model for the distributions of $\Am$ and $\ra$, based on three abstractions which should be robust to the spatial bias of the data.
Nonetheless, we point out that, as the dataset is oriented towards paths incorporating long, symmetric inter-continental links, the relative asymmetries may be considerably smaller than when paths are all within a single region. Conversely, the absolute values of asymmetry may be more representative than the large RTTs may suggest.  We restrict to a study of asymmetry distribution because, as the data is all from a single client viewpoint, it cannot, unfortunately, be meaningfully used to study asymmetry dependencies across the network topology.  Despite these limitations, it is worth bearing in mind that the data displayed in Figure~\ref{fig:scatter} constitutes {\dv by far} the most authoritative collection of precision path asymmetry information yet collected for the Internet, and the model below is the first we are aware of based on reliable data.

\begin{figure}[b!]
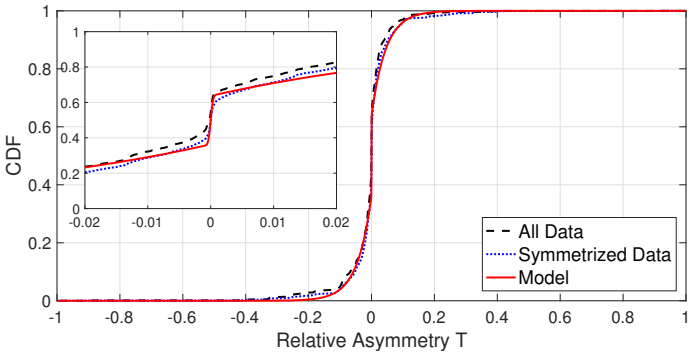

    \begin{center}
     \hspace*{-0.1cm}  \RACDFandmodel  \hfill
    \end{center}
    \vspace{2mm}
    \caption{CDF of the measured relative asymmetry $\era$, its symmetrized version, and fitted mixture model.}
    \label{fig:RACDF}
\end{figure}

The first abstraction is that of symmetry. As argued above, there are structural reasons why $\Am$ and $\ra$ should be symmetrically distributed about zero.
We therefore fit symmetric distribution families to the symmetrized form of the data, also plotted in Figure~\ref{fig:RACDF}.
That is, if $\{y_i\}$ is the set of $\era$ values, we fit to the CDF of $\{y_i\} \cup  \{-y_i\}$.

The second abstraction is the approximate independence of relative asymmetry on RTT, leading to a model where $\Am = \Rm\, \Tr$, where $\Tr$ is a symmetric random variable on $[-1,1]$, 
\textit{independent of $\Rm$}.   This implies that samples of asymmetry can be generated simply by scaling up a random variable $\Tr$ by a deterministic RTT, an appealing feature.

The third abstraction is building in the peak. Inspection of the symmetrized data indicates its support has width of the order of only $w=0.001$, despite containing over 20\% of the probability. 
This is too pronounced even for distribution families such as the Generalized Gaussian distribution which have a shape parameter regulating peakedness. 

\smallskip
Based on the above, we propose a mixture model  combining a centred uniform distribution $U$ of support width $w$ to account for the peak,  and a centred Laplace variate $L$  renormalized over $[-1,1]$.  A Laplace distribution is itself peaked, fits the tails adequately, and is simple.
The mixture variable selects $U$ with probability $p$, else $L$.
We use $w=0.00136$, select $L$'s scale parameter $b$ based on the Maximum Likelihood estimator $\hat b = \sum_{j=1}^{m} {|\era_j|}/m = 0.0450$
 for the Laplace component, and numerically solve for the $p$ which minimizes the $L1$ distance to the symmetrized data over the quartile range $[q_{0.2},q_{0.8}]$, yielding $p=0.274$.  The CDF of the model compares well with the symmetrized data in Figure~\ref{fig:RACDF} {\dv over the central region covering 98\% of the probability}.

 From Figure~\ref{fig:RACDF},  $\era$ lies within $[-0.02,0.02]$ in the symmetrized data 60\% of the time, corresponding to $|E/\Rm|<0.01$.  This suggests, {\dv soberingly,} that to achieve errors below (1\mus,10\mus,1\ms) for 60\% of paths, $\Rm$ values not exceeding (100\mus,1\ms,100\ms) will be required.
 Using the model, we can infer errors in a similar way for RTT values not well represented in the data.


\subsection{The Danger: Asymmetry Jitter}
\label{ssec:jitter}

Client clocks, in particular the ntpd daemon, often consider a selection of servers in parallel, and may switch between them as conditions change. In addition, services like \textit{ntppool} \cite{ntppoolproject} use DNS to change the server accessed by the client via a constant url.   
Consider now the impact from an asymmetry perspective:
each change in server implies a change in path, and hence a change $\Delta(\Am)$ in asymmetry, and $\Delta(E)=\Delta(\Am)/2$ in the clock error. 
Moreover, even if a change is favorable in the sense of a server with a lower underlying asymmetry $\Am$, the sudden change itself is not, causing for example jumps in OWDs which may confuse software, and even disrupt the clock synchronisation algorithm.  We refer to this unwelcome movement as \textit{asymmetry jitter}. 
We now use the data and model to gain insight into the magnitude of this problem.

\begin{figure}[b]
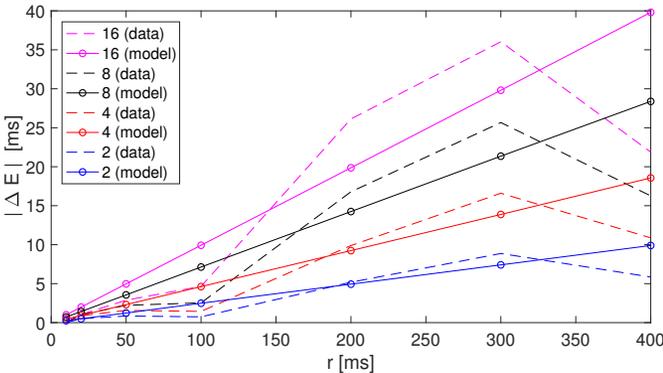

    \begin{center}
        \hspace*{-0.5cm}\jitterError
    \end{center}
    \vspace{1mm}
    \caption{Range of asymmetry jitter among groups of $\{2,4,8,16\}$ servers as a function of $\Rm$.}
    \label{fig:jitterError}
\end{figure}

Consider a scenario where a client may move among a set of $N_s$ servers, selected from some pool. The range of variation in $\Delta(\Am)$ over all pairs of servers from the pool is simply given by the range (i.e.~$\max-\min$) of the $\Am$ values themselves.
In Figure~\ref{fig:jitterError} we plot the corresponding clock error for pools defined by $\Rm$ `annuli', given by $\Rm\in[0.8r^*,1.2r^*]$, where the central $r^*$ values are shown in the figure.  
Results are shown for server sets of size $N_s=\{2,4,8,16\}$.  We also compare against the model, where $\Am$ values are randomly drawn corresponding to $\Rm$ values chosen uniformly in the same range as for the data. 
Both for the data and the model, 100,000 replications were used for each point to accurately estimate the average over all possible selections from the pool/ensemble.
The data shows a rough agreement with the model where data is sufficient, and the increase with $\Rm$ is another validation of the latter's $\ra$-based structure.  
When the pool has weak diversity, in particular at $r^*=\{10,100,400\}$\ms, {\dv the model and data disagree substantially. This should not be seen as weakness in the former. On the contrary, it illustrates the usefulness of the model in interpolating over gaps in the data.}



The main observations on jitter errors are that
(i) the problem gets worse with more servers:  there is no averaging effect here, more servers represent more opportunity for jitter,
(ii) the increase with RTT highlights the importance of limiting the damage through selecting lower $\Rm$,
(iii) jitter can reach damaging values, despite the high concentration of relative asymmetry about zero.

\subsection{The Opportunity: Bound Tightening}
\label{ssec:bounds}

We have just seen that synchronizing to alternative servers risks a penalty due to asymmetry jitter. However, it turns out that multiple servers can be used in another way to gain a benefit instead: tighter bounds on asymmetry. 

To describe this concisely, we adopt an `interval calculus' notation, where 
$b+ [c,d]$ denotes the interval $[b+c,{\dv b+d}]$, $-[c,d] = [-d,-c]$, and an overbar denotes  an interval
variable.   For example we can write the default \textit{asymmetry interval}, {\dv by which we mean the smallest interval} within which $\Am$ must lie, as $\widebar\Am= [-\Rm,\Rm]$. 
Recall that $E_a = (\Am-a)/2$.  Using  $\widebar\Am$, it is easy to derive
the \textit{error interval}, within which $E_a$ must lie, as 
$\widebar E_a = (\widebar \Am - a)/2$,  which has width $\Rm$ and is centered at $-a/2$ (this is equivalent to the more cumbersome $-\Rm/2 -a/2\le E_a\le \Rm/2-a/2$).

Now consider the \textit{clock interval}, by which we mean the smallest interval, computable by the clock itself, within which the true time $t$ is known to lie. Since $t=C_a(t)-E_a$, the clock interval (which is independent of $a$) is 
\be
   \label{eq:clockinterval}
   \tb = C_a(t) - \widebar E_a = C_0(t) - \widebar E_0 = (t + \Am/2) -\widebar \Am/2 \ .
\ee

To distinguish between servers we now add a superscript $j\in\{1,2,3\ldots\}$.  Hence $\Rm^2$ is the RTT from the client to server 2, $\Am^1\in[-\Rm^1,\Rm^1]$ is the true asymmetry from the client to server 1 and its initial bound, etc.  Thus at the client there is a separate instance of the clock model for each server $j$, complete with its own asymmetry interval $\widebar \Am^j$ and parameter $a^j$.

To show that sharing information across multiple servers can help, consider a trivial case with two servers where we assume $\Am^1$ is known. Clock 1 then sets $a^1=\Am^1$ and achieves zero error.  Clock 1 can then be used to measure the true OWDs and hence the asymmetry $\Am^2$ to server 2. This information is passed to clock 2, who sets $a^2=\Am^2$ and becomes itself perfect.   
A second example:  assume $r=\Rm^1=\Rm^2$ and $\Am^1=r=-\Am^2$, with $a^1=a^2=0$.    Imagine each clock timestamps some event which occurs at time $t^*$, which we can take to be $t^*=0$.  
Clock 1 reads $C_0^1(0) = r/2$ and knows that the true time must have occurred inside its clock interval $[0,r]$. Similarly clock 2 reads $C_0^2(0)=-r/2$ and knows that the true time must lie in $[-r,0]$. The clock intervals intersect at a single point, namely $0$, which must therefore be the true time. 
Using this fact, each clock can infer its asymmetry exactly, and reset its asymmetry configuration parameter $a^j$  to it. 

We now extend the above to $N$ servers, and 
consider the general case where the client knows a valid initial asymmetry interval $\widebar \Am^j=[\al^j,\ar^j]$ to each server $j$, which must lie within the default interval $[-\Rm^j,\Rm^j]$.
Our goal is to show how the set $\{\widebar \Am^j, j=1\ldots N\}$ of initial bounds can be combined to produce tighter updated bounds for each server.
As before,
without loss of generality we will compare clock readings at $t^*=0$.

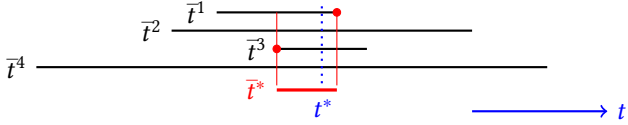
\begin{figure}[tbh]
\hspace{0mm}
\vspace{-2mm}
\centering
%
%
%

\begin{tikzpicture}[scale=2]       

\draw[-, thick]  (1.3,0.76) node[left] {$\widebar t^1$} 	-- (2.1,0.76);
\draw[-, thick]  (1.0,0.64) node[left] {$\widebar t^2$}  	-- (3.0,0.64);
\draw[-, thick]  (1.7,0.52) node[left] {$\widebar t^3$}  	-- (2.3,0.52);
\draw[-, thick]  (0.1,0.40) node[left] {$\widebar t^4$} 		-- (3.5,0.40);

\draw[-, red]  (1.7,0.76) 	 			-- (1.7,0.28) ;    
\draw[-, red]  (2.1,0.76) 	 			-- (2.1,0.28) ;	
\draw[-, blue,dotted,thick]  (2.0,0.80) 		-- (2.0,0.28) ;	

\node[draw,circle,inner sep=1pt,fill,red] at (2.1, 0.76) {};     
\node[draw,circle,inner sep=1pt,fill,red] at (1.7, 0.52) {};     

\draw[-, very thick, red]  (1.7,0.25)  node[left] {\!$\tb^*$}  -- (2.1,0.25);
\node at (2.01,0.13) {$\color{blue} t^*$};

\draw[->, thick,blue]  (3.0,0.11) 	-- (3.9,0.11) node[right] {$t$};

\end{tikzpicture}
    \vspace{2mm}
    \caption{Intersecting 4 clock intervals to form a tighter $\tb^*$. The true time, $t^*\in\tb^*$, is unknowable to clients.  
  }
    \label{fig:reconciled}
    \smallskip
\end{figure}

It is convenient to denote the clock interval of $C_{a^j}^j(0)$ as $\tb^j = [\cl^j,\crr^j]=(\Am^j-\widebar \Am^j)/2$.
The key observation is that, since the true time $t=0$ must lie in each of the time intervals, it must lie in their intersection, the \textit{reconciled time interval}, which 
can be written simply as
\be
\ntinyskip
   \label{eq:Istar}
   \tb^* = [\cl^*,\crr^*] = [\max_j \cl^j, \, \min_j \crr^j] \ ,
\ntinyskip
\ee
and which becomes the updated time interval for \textit{all} clocks (see Figure~\ref{fig:reconciled}).
For each $j$  the updated $\tb^j=\tb^*$ may be unchanged, or truncated at either or both sides compared to the original.  
The original intervals with small $\Rm^j$ will tend to control the intersection, however those with larger $\Rm^j$ can also be of influence if their asymmetry $\Am^j$ is also large.
Although the asymmetry intervals and hence the reconciled interval are known to the client, 
the position of $t^*$ within them, though shown in Figure~\ref{fig:reconciled}, is not.

With $\tb^*$ known, each clock can update its bound as
\ba
  \label{eq:newAbar}
   \widebar \Am^j = 2C_{a^{j}}^j(0) + a^j - 2[\cl^*,\crr^*]
                                    = 2C_{0}^j(0)              - 2[\cl^*,\crr^*] ,
\ea
and then benefit from it by updating $a^j$ to lie within it, the natural choice being its midpoint.

\begin{figure}[t]
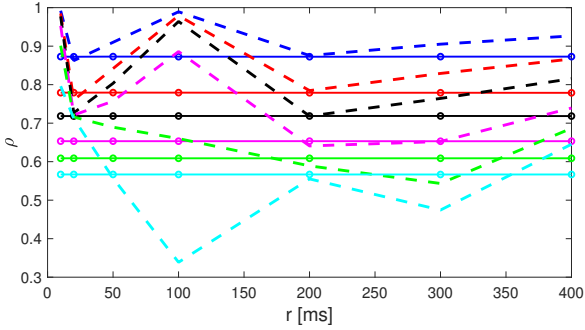

\nsmallskip
    \begin{center}
\hspace*{-0.5cm}  \boundtightening
    \end{center}
    \vspace{1mm}
    \caption{Degree of bound tightening among groups of (in order from top to bottom) $(2,4,8,32,128,512)$ servers as a function of $\Rm$, for data (dashed lines) and model.}
    \label{fig:boundtightening}
\end{figure}

Following an analogous procedure to that of Section~\ref{ssec:jitter}, we can use the data, and model from Section~\ref{ssec:model}, to evaluate the degree of expected tightening based on server sets of size $N_s$, and initial default asymmetry intervals $[-\Rm^j,\Rm^j]$. This will be expressed through the ratio 
 $\rho = | \tb^* |/\max_j(\Rm^j)$,  
of the reconciled width to the loosest initial width across the set.
Figure~\ref{fig:boundtightening} displays the expected value of $\rho$ based on 100,000 server set selections, using the same $\Rm$ annuli as above, for each of $N_s=\{2,4,8,32,128,512\}$.

The results for the model are independent of $\Rm$ by construction, and the data shows a similar pattern except where the pool of $\Rm$ in an annulus is insufficient, as noted above.

The main observations for the tightening are that
(i) the situation \textit{improves} with more servers: each new server represents a possible further tightening,
(ii) the degree of tightening is modest, large numbers of servers are needed to reach a significant reduction, indicating insufficient diversity of relative asymmetry for large gains in typical cases,
(iii) the approximate independence w.r.t.~the minimum RTT $\Rm$ implies that tightening of asymmetry $\Am$ itself will be proportional to $1/\Rm$, and so it \textit{will} become significant, if $\Rm$ is low enough.  Thus, $\rho$ being modest again highlights the importance of asymmetry error mitigation through the selection of servers with low $\Rm$.

\section{Changing the Rules}
\label{sec:meth}

The previous section described how the limits to clock error set by asymmetry are fundamental.
How then can they be circumvented?   The answer can only be by exploiting independent, a priori information about the network. We now describe two incarnations of this idea.

\subsection{A Deterministic Latency Path Model}
\label{sec:bounds}

The default bound $\Am\in[-\Rm,\Rm]$ arises from the simple causality bounds for OWDs, namely $\dfm\ge0$, $\dbm\ge0$.  However, we know from the physics of networks that, for any given round-trip path,  $\dfm\ge\dfb$ and $\dbm\ge\dbb$ for some suitable lower bounds $\dfb$, $\dbb$.
The  asymmetry interval then tightens to 
$[2\dfb-\Rm,\,\Rm-2\dbb]$, 
and the corresponding error interval $\widebar{E}_a= [\dfb-\Rm/2,\,\Rm/2-\dbb]+a/2$ has width reduced  
to $\Rm - (\dfb+\dbb)$.
Tighter bounds on individual paths translate directly to tighter bounds when exploiting multiple servers using (\ref{eq:newAbar}), exactly as before.

A simple but reliable way to obtain such path bounds is to use a Speed of Light (SoL) approximation coupled to a lower bound on path distance.  Specifically, we use values based on the SoL of $2c/3$ (based on the refractive index of common fibre of $1.5 = (2/3)^{-1}$),
over the great circle distance between client and server, assuming locations known by geolocation or otherwise, resulting in $\dfb=\dbb>0$.

\miniskip\noindent\textbf{Algorithm: SoL Based Bound Estimation (SBBE)}  \\ 
In practice a client clock must work with a finite number of exchanges with the server, and the minimal path characteristics are unavailable. 
Instead the measured OWDs follow $\df = \dfm +\qf$,  $\db = \dbm + \qb$, and
$R=\df+\db=\Rm+q$, where $\qf\ge0$, $\qb\ge0$, and $q = \qf+\qb\ge0$ represent congestion variability.
To attempt to approximate the zero variability limit the algorithm proceeds as follows. 
The $i$-th exchange gives rise to an error interval of width $R_i - (\dfb+\dbb)$, where 
the position of the interval can be calculated from available timestamps as 
$[e_l^i,e_r^i]=[C(t_a)-t_s+\dfb,\, C(f_f)-t_s-\dbb]$.  
The \textit{refined} error interval that satisfies all these constraints is then 
\be
  \widebar{E}^* = [e_l^*,e_r^*]=[\max_i e_l^i,\min_i e_r^i] .
  \label{eq:CBOE}
\ee
Now, the refined intervals using other servers, using the same client clock but each with their own $\dfb$, $\dbb$, can be combined in the same way to produce a final bound 
(this single clock view is equivalent to the multiple clock view of equation~(\ref{eq:newAbar}). Think of $C^1_0$ as the nominal client clock, with the other clocks aligned to it via implicit $a^j$ choices), 
and we define the associated offset error to be its central  value.
We term this approach \textit{SoL Based Bound Estimation (SBBE)}.

To summarize, SBBE is based on a client collecting round-trip data to a set of servers.  Each exchange gives rise to an error interval (within which the true error must lie) which benefits from a simple SoL-based bound for each OWD.  Intervals are  combined `within' servers to combat path variability, and across servers to exploit asymmetry and $\Rm$ diversity (as well as more data). 
The end result is a much tighter interval, and a more accurate estimate of offset error.



\miniskip
\miniskip\noindent\textbf{\dv Note on Interpretation:} \quad   The offset error above should be thought of as an assessment of the systematic error component of the client clock, which is not itself changed/corrected.
Accordingly, the offset measurement proposed here should \textbf{not} be confused with a dynamic clock synchronization algorithm, which would run the risk of the asymmetry jitter described in Section~\ref{ssec:jitter} if choosing to move between servers. 
(Notwithstanding the fact that this error, although primarily about asymmetry,
may also include the correction of a systematic constant error, the $\theta$ from (\ref{eq:offsetmodel}), due to the algorithm itself.) 
Here multiple servers are involved only for the purpose of asymmetry bound reduction. 
We do not enter into the subject of how the client clock is dynamically synchronized, as this is a distinct problem requiring a paper to itself.  For concreteness however, it is best to imagine that only a single nearby server, nominally server $j=1$, using $a^1=0$, is used for this purpose.
The same comment holds for the method we describe next, 
{\dv and throughout the remainder of the paper}.


%
%
%

\vspace{-0.2cm}
\subsection{A Statistical Latency Network Model}
\label{sec:calib}

Bounds on OWDs and asymmetry could also be obtained through knowledge of the `latency topology' of the network, for example
through establishing a mapping between latency and network `position', linked to geographic position.
Such bounds will be statistical in nature, estimates rather than the true bounds of the SBBE approach above, resulting in a tradeoff of tightness against accuracy.

We consider a scheme where a set of $L$ landmark servers with synchronized clocks (nominally, participating stratum-1 servers), and known locations, exchange error-free OWD measurements with each other, in order to establish a  mapping function linking geographic location to OWD network bounds. This mapping  is `universal' in that it is not a function of explicit network structure. Instead it is learnt based on sample OWDs, and is taken to be representative of the inter-landmark network and its environs.
It will not apply strictly to any client, as the inter-landmark and landmark$\leftrightarrow$client paths differ.  On the other hand by exploiting the network diversity as sampled by landmarks, multiple approximate bounds for a client may be tightened and `cross checked'. 
As with SBBE, the geographic model will be great circle distances, calculated assuming geolocation information is available, and like SBBE, the corresponding SoL bounds will be incorporated.

When a client exchanges timestamps with a landmark,  bounds are calculated by the landmark based on a location provided by the client, and communicated back to it with the returning timestamp.  Thus each timestamp exchange is also a query for bound estimates for each direction. 

\begin{figure}[htb!]
  \centering
    \epsfig{figure=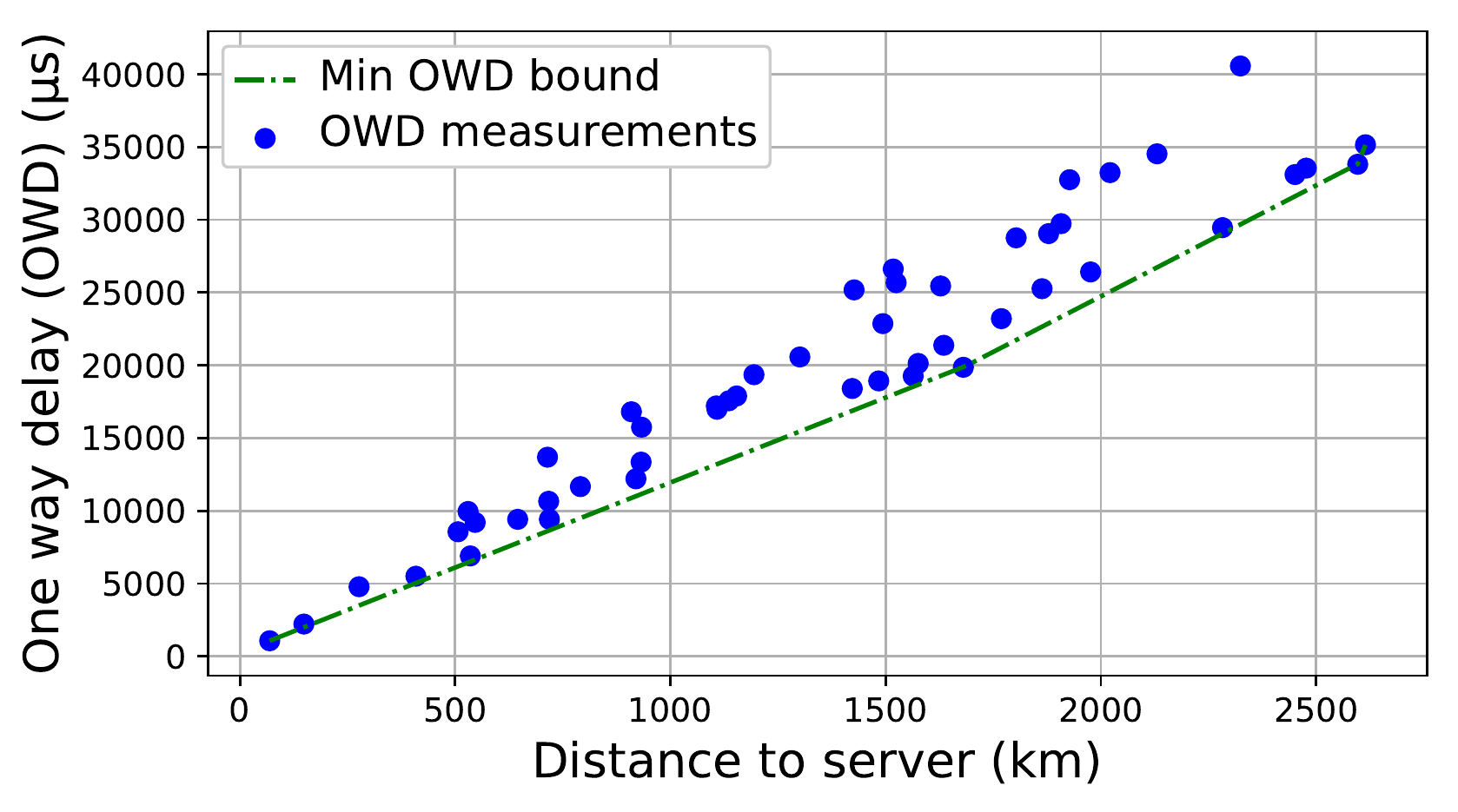,width=8cm}
    \miniskip
 \caption{Example landmark mapping function.}
 \label{fig:landmark_calibration}
\end{figure}

\miniskip\noindent\textbf{Algorithm: Landmark Based Bound Estimation (LBBE)}\\  
Each landmark establishes two independent mapping functions, one for each OWD direction, based on its own round-trip exchanges to other landmarks. 
Inspired by the approach of \cite{Wong07}, each mapping function is generated as follows. 
A scatterplot of the OWDs against the (great circle) distances is first plotted, and a convex hull calculated around it.
The lower facet of this hull (green curve in Figure~\ref{fig:landmark_calibration}), is a convex function $M$ mapping any distance $x$ within the observed range to 
an estimated lower bound $M(x)$ on delay.  
When queried by a client, for each map the landmark returns an estimated bound of $\max\{M(x),d_L\}$, where $d_L$ is the client-specific SoL bound for that direction.  
So that the landmark estimate is never based on insufficent data, 
it defaults to $d_L$ if $x$ falls outside the $[0.2,0.8]$ quantile interval of available $x$ values.  

With two caveats, the remainder of the algorithm is now as for SBBE:  multiple error intervals arising from multiple queries of each of some subset of landmarks are combined following equation~(\ref{eq:CBOE}), simply the bounds $(\dfb,\dbb)$ are replaced with those from the landmarks, for which generally $\dfb\ne\dbb$.
The caveats are (i):  bounds can now vary in time as the mappings update; (ii) as the bounds are statistical, they may be inconsistent, with a null intersection.
In that case, we process the error intervals according to a heuristic that, essentially, moves from tighter to looser bounds until the first consistent choice.  
Finally, to ensure the correctness of the error constraints provided to clients, it is important to ensure the statistical validity of the mapping functions. To this end, {\dv in practice each landmark server must re-calibrate} itself periodically.

To summarize, LBBE uses OWD bounds provided from a geographic distance to OWD bound predictor, established empirically from OWD measurements between a set of landmark servers.   It is a statistical approach which may learn more about the network latency topology, resulting in tighter bounds, at the cost of inevitable errors in the mapping as applied to client locations.

\section{Evaluation}
\label{sec:evaluation}
\newcommand{\dsol}{D_{{\tiny SoL}}}

The goal of this section is to gain insight into the comparative performance of SBBE and LBBE, and how each method behaves as a function of key parameters.
As a full evaluation, for LBBE in particular, would require access and control over a large set of geographically distributed stratum-1 servers, we turn to simulation instead.  We use unstructured network topologies with no asymmetry dependencies.  This represents the hardest environment for LBBE to operate in, where the bound information it passes to clients is effectively reduced to that of the asymmetry marginal distribution only.

\vspace{-0.2cm}
\subsection{Simulation framework}
\label{ssec:simulation}

In keeping with the approach of the paper, we target our evaluation at the level of fundamentals, rather than fully realistic network assumptions. 
The simulation therefore follows the clock and path models established in Section~\ref{ssec:fund}. The details are as follows.


\miniskip\noindent\textbf{Servers}  \quad 
All server clocks are perfect, and there is zero delay between the reception of a message from a client on the forward path and its retransmission backward {\dv (i.e.~$t_s = t_i = t_o$ from Section~\ref{sec:asym}).}

\miniskip\noindent\textbf{Client clocks}  \quad 
Client clocks are taken to be perfect except for a constant offset error $E$, which is chosen uniformly at random over $[-10,10]$\ms. 
These ground truth errors are generated once only then held fixed.
The methods will estimate these errors, but not correct them.

\miniskip\noindent\textbf{Server and client placement}  \quad 
We place 50 servers, and 1000 clients, randomly over a region, defined by latitude and longitude, approximating the geographic extent of the mainland US. Locations are generated once only then held fixed.
All servers are used as landmarks.

\begin{figure}[tb!]
  \centering
 	 \epsfig{figure=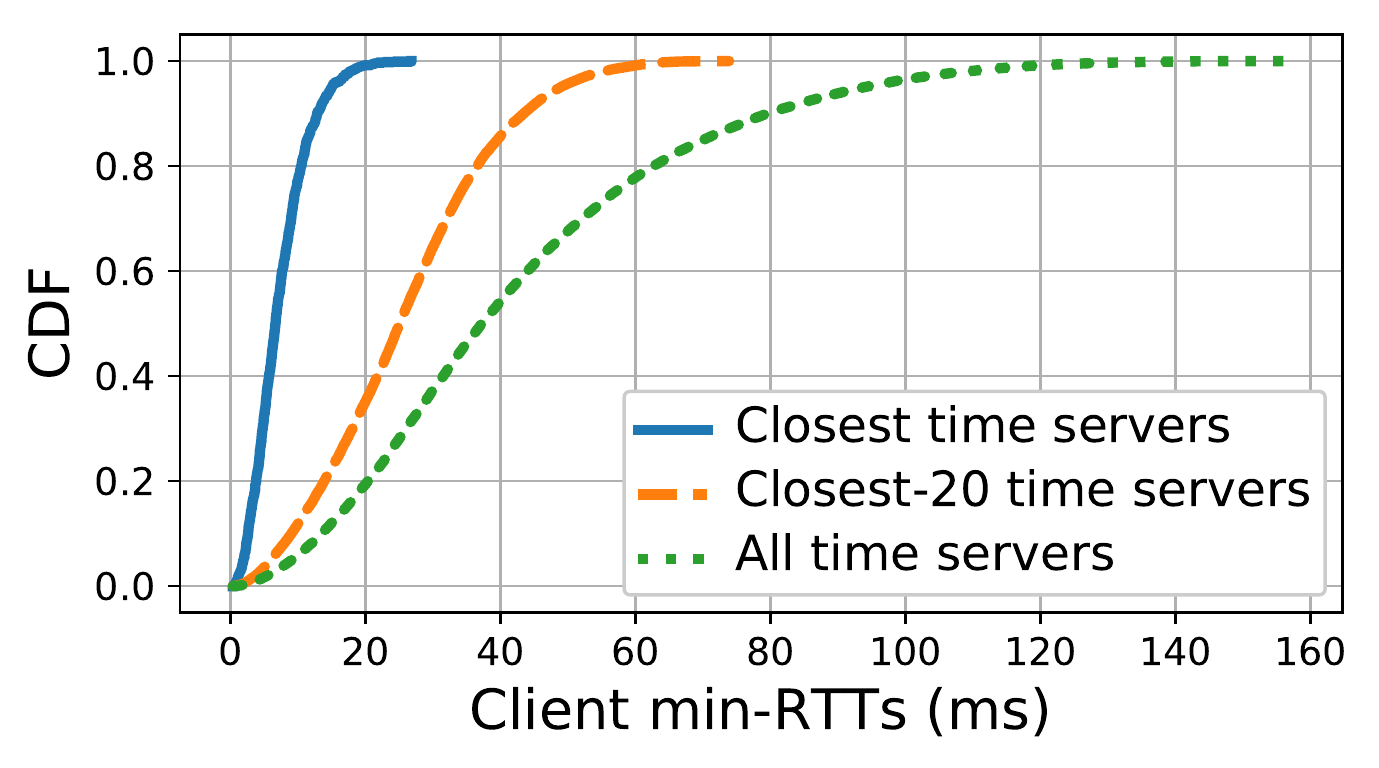,width=8cm}
 	 \miniskip
 \caption{ CDF of client $\Rm$ values used in the simulation.}
 \label{fig:mRTTCDF}
\end{figure}

\miniskip\noindent\textbf{Underlying path parameters}  \quad 
We first use our geographic based network model to generate the minimum RTT  between all (server, client) or (server, server) pairs, randomly from their locations. 
Specifically, for each pair,  $\Rm$ is selected as
$\Rm=F\cdot2\dsol$, where the inFlation factor $F$ is uniformly distributed over $[1.2,1.8]$, and 
$\dsol$ is the SoL in fibre delay over the great circle distance between the pair.  Here $F$ accounts for the error between the great circle model and the underlying network.
The CDF of all generated $\Rm$ values in the simulation is given in Figure~\ref{fig:mRTTCDF}.
The asymmetry between the pair (in {\dv a random direction for each pair}) is then generated as $\Am = \Rm\, \Tr$ using the model of Section~\ref{ssec:model} (renormalized to an interval $[-T_L,T_L]\subset[-1,1]$, see below, to respect the SoL bounds $\dfb$, $\dbb$).  
This is negated in the other direction in the case of two landmarks when they probe each other.
Each $\Rm$ and $\Am$ are generated once only then held fixed.
They then define the minimum OWDs: $\dfm = (\Rm+\Am)/2$, and $\dbm=(\Rm-\Am)/2$.

\miniskip\noindent\textbf{Congestion model}  \quad 
Over any OWD, the variable components $\qf$ or $\qb$ of $\df$ and $\db$ respectively, as described under SBBE in Section~\ref{sec:meth}, are simply taken to be independently and identically distributed according to an Exponential distribution with default mean $\mu=1$\ms.
Richer models incorporating temporal correlations, and path-dependent congestion levels, would result in more complex behavior and in general worse performance. A simple model is sufficient for our purposes here, being focussed on asymmetry fundamentals. {\dv In fact, in the results below we often restrict to the limiting case of zero congestion to isolate the core insights, and the best case potential of the methods.}

\subsection{Sensitivity analysis}
\label{ssec:sensitivity}

With the client and server positions, and $\Rm$ values, set as described above, we now examine performance as a function of three main variables:  server choice, path congestion, and asymmetry diversity.  Throughout, results are given averaged over the 1000 clients, and where applicable, 
over the 100 independent replications of the congestion random variables. 





\begin{figure}[b!]
 \centering
  	\epsfig{figure=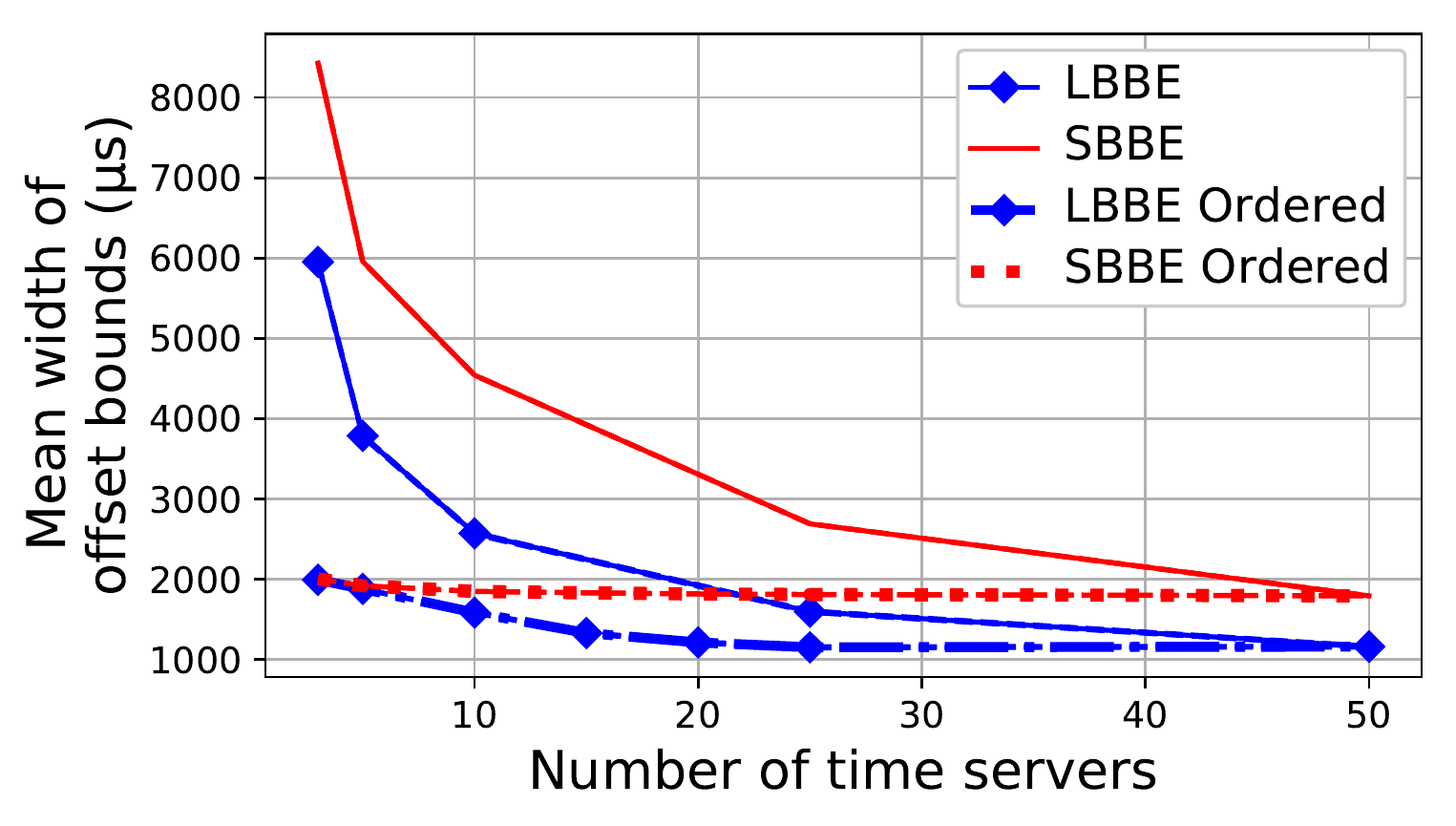,  width=69mm} \ \,
  	\epsfig{figure=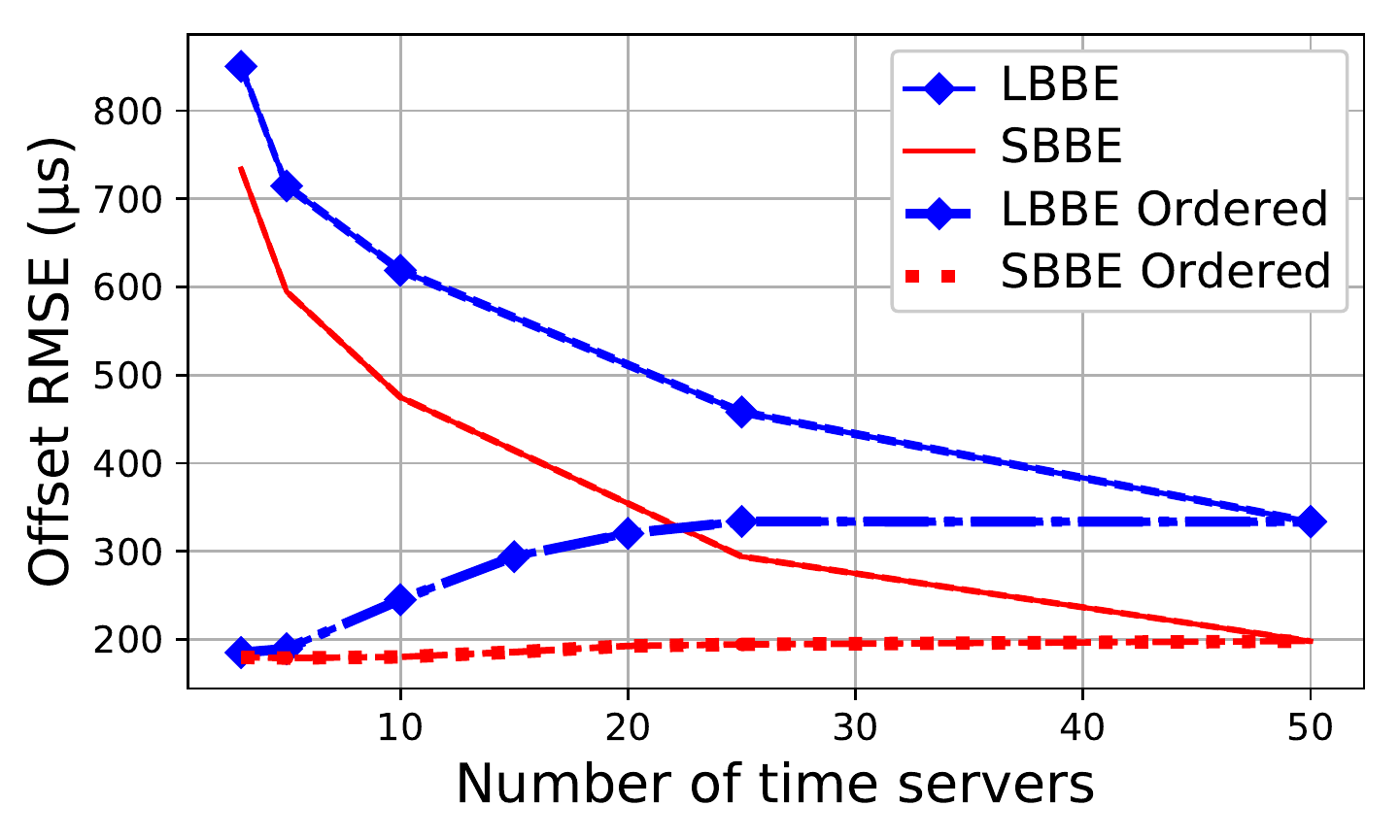,width=65mm}
  	\smallskip
 	\caption{\dv Left: offset bound width as a function of \# servers, 
 	        Right: RMSE of the corresponding $E$ estimates.}
 	\label{fig:numLandmarks_bounds}
\end{figure}

\miniskip\noindent\textbf{Time server choice / Role of $\Rm$.}  \ We first consider the crucial role of the number, and placement, of the servers used by clients.
For this purpose we select a modest level of congestion of $\mu=1$\ms, and use $M=16$ measurements
to each server. This value is sufficient {\dv for this $\mu$}, as justified below.  


\begin{figure*}[t]
 \centering
 \epsfig{figure=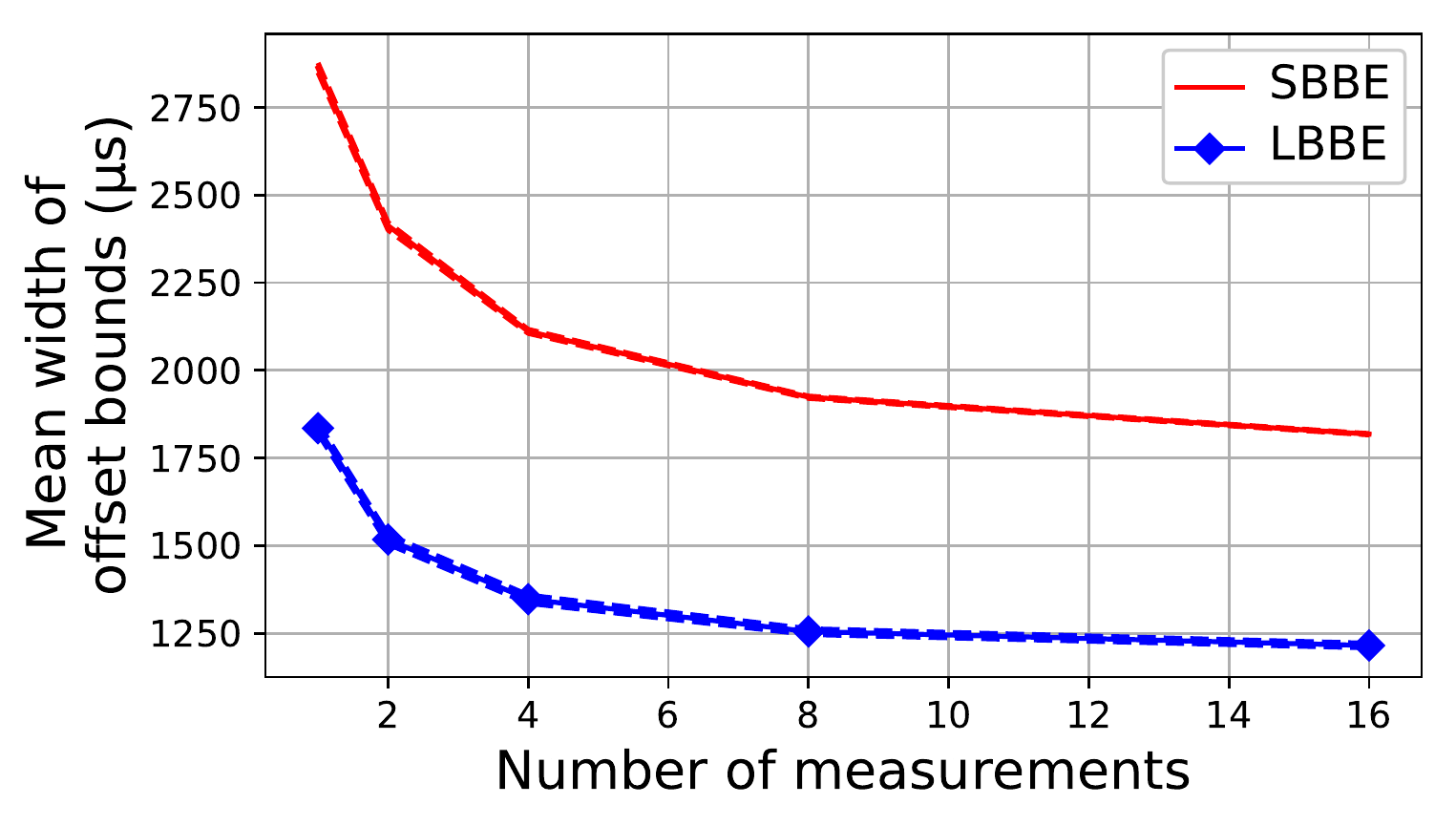,width=69mm} \ \,
 \epsfig{figure=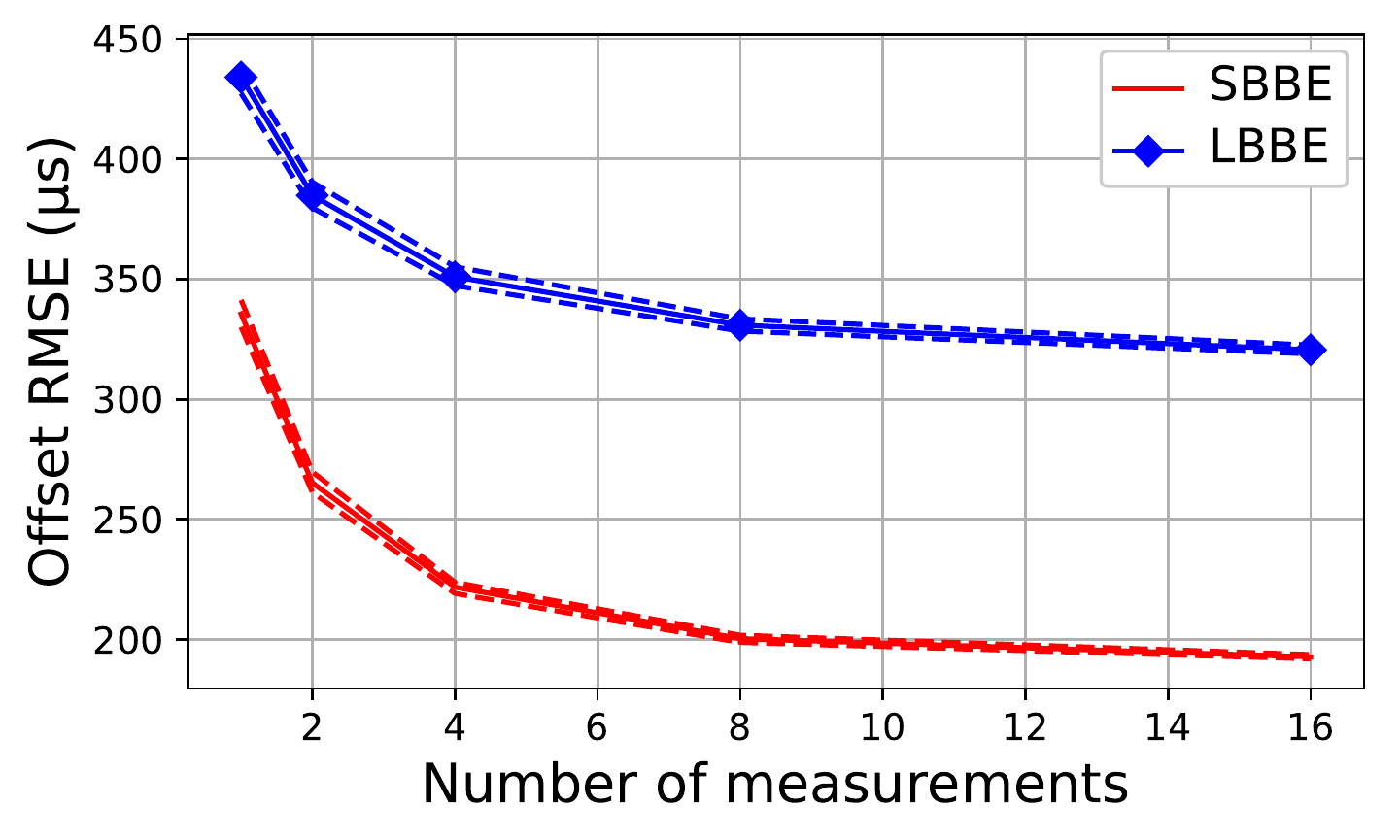,  width=65mm}
 \vspace{0.2cm}
 \caption{\dv Bound width (left) and RMSE (right) as a function of M.}
 \label{fig:bounds_rmse_icbeNumMeasurements}
\end{figure*}


Figure~\ref{fig:numLandmarks_bounds} gives the resulting bound widths as a function of the number of servers used by each client.
In all cases, bounds tighten as information from more servers is included, as expected. 
In the naive case (solid lines) servers are selected randomly: errors start high and many servers are needed in order to capture those allowing bound tightening.  
In the Ordered cases (dashed lines), each client uses the closest servers to it:  the bounds start low and move even lower, saturating after only a few servers, 10 being sufficient for SBBE, and 20 for LBBE.
These results confirm the important role of the closest servers, and are consistent with the 
information on the `supply' of $\Rm$ available to clients from Figure~\ref{fig:mRTTCDF}, 
linked to the predictions at the end of Section~\ref{ssec:model}.

The saturated bound width from {\dv the left plot in} Figure~\ref{fig:numLandmarks_bounds} of LBBE is almost half that of SBBE, however when we look 
at the RMSE (over clients) of the point estimates of $E$
{\dv in the right plot}, 
namely the bound midpoints, we find SBBE is the better choice.  In fact LBBE errors actually increase with the number of servers!  We explain the reason for this presently.  
%
 
To complete this section on server selection and the role of $\Rm$, we increase server density by a factor of 100 by scaling geographic distances down by a factor of 10.  In this shrunken world,  servers are 10 times closer to clients, with the median $\Rm$ to the closest server dropping from $6.7$\ms to $0.9$\ms.
Using the 20 closest servers only, the RMSE errors for (SBBE, LBBE) become (26,38)\mus, 
almost a tenth of those previously.
 

\miniskip\noindent\textbf{Path congestion / Number of measurements.} \   
The number $M$ of timestamp exchanges performed with each synchronizing server influences accuracy through combating congestion variability.  
Again using the 20 closest servers only, and $\mu=1$\ms, {\dv Figure~\ref{fig:bounds_rmse_icbeNumMeasurements} shows bound width and estimation error both}
decreasing rapidly as $M$ increases, but with little change after $M=16$.





\def\bb{\!\!\!}  
\def\bb{} 
\begin{table}[b!]
\centering
    \scalebox{0.9}{
    \begin{tabular}{ c|c|c|c|c|c|c }
    Congestion  ($\mu$) & 0\ms & 0.5\ms & \textbf{1\,ms} & 5\ms  & 10\ms  &  50\ms\\ \hline
    SBBE & 190 & 190 & \textbf{193} & 232 & 286 & 522\\
    LBBE & 311 & 316 & \textbf{363} & 350 & 411 & 655 \\
    \end{tabular}
}
\vspace{4mm}
    \caption{ RMSE (\mus) at different congestion levels, \dv using  $M=16$ measurements.}
    \label{tab:rmseCongestion}
\end{table}
Table~\ref{tab:rmseCongestion} (still with $M=16$) gives an idea of how errors in $E$ estimation 
vary as a function of congestion. 
It includes the special case of no congestion, $\mu=0$, which reveals the underlying limitations of the available bound diversity. 

%

\begin{figure}[t!]
  	\centering
\epsfig{figure=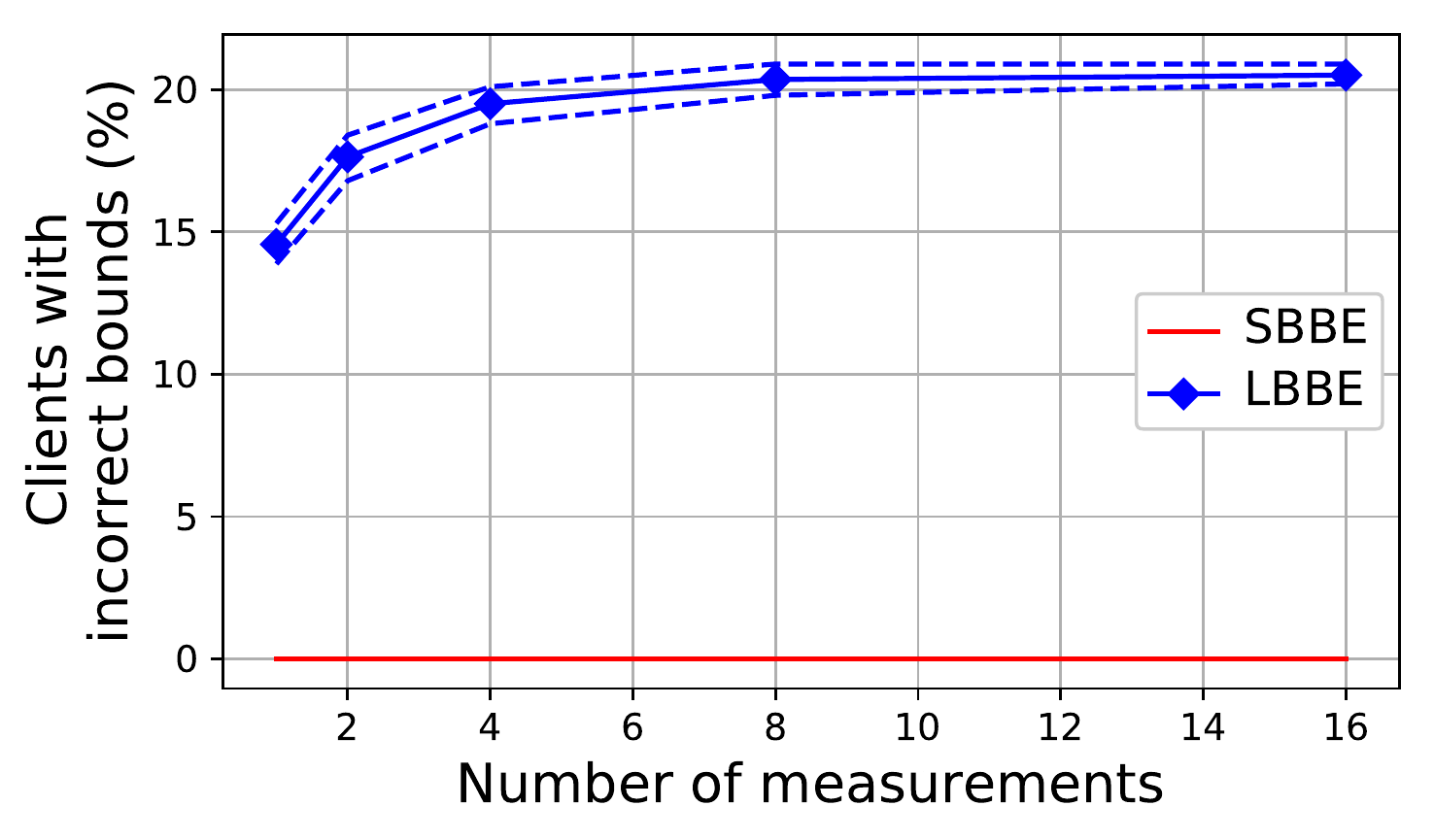, width=65mm} \ \, 
   \epsfig{figure=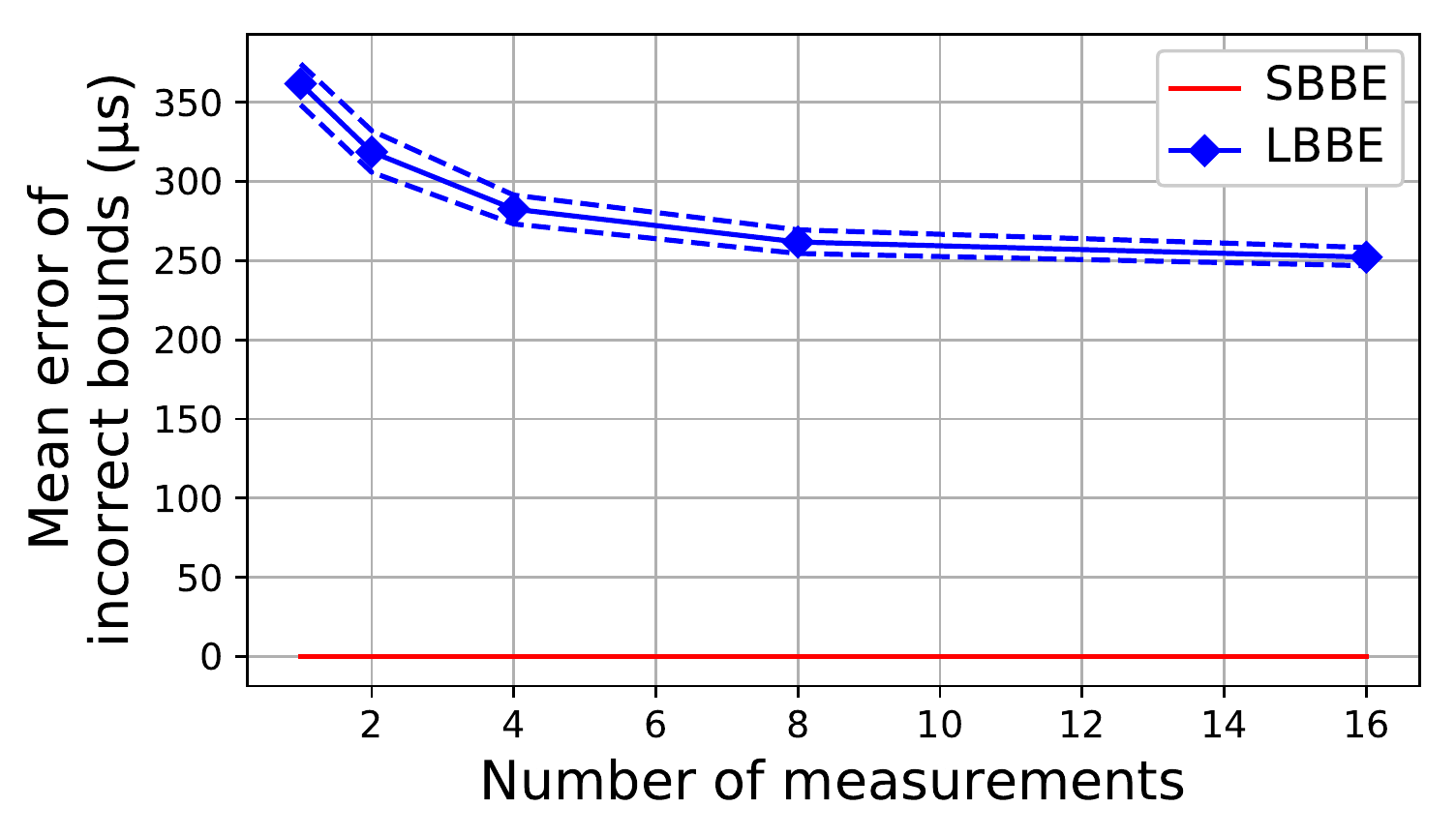,width=67mm}
\miniskip
 \caption{\dv Tracking inconsistent bounds. Left: percentage of clients with an inconsistent bound, Right: average error of those bounds that are incorrect.}
 \label{fig:incorrectBounds}
\end{figure}

The difference in underlying nature between the SBBE and LBBE methods is on display in Figure~\ref{fig:incorrectBounds}. 
{\dv Consider the left hand plot.}
Since SBBE is deterministic with provably correct bounds, the number of clients that receive an inconsistent final bound, that is, the true value of $E$ lies outside of it, is zero, whereas for LBBE around 20\% of clients receive an inconsistent bound.
{\dv  The right plot in Figure~\ref{fig:incorrectBounds}}
measures how far away the closest edge of these incorrect bounding intervals are from $E$, on average. 
Although, unlike the client percentage, this distance decreases somewhat with $M$, it remains significant, around 250\mus at $M=16$ compared to a total error of 360\mus 
(Figure~\ref{fig:numLandmarks_bounds}).
We can now explain the apparent contradiction between 
{\dv the plots in Figure~\ref{fig:numLandmarks_bounds}}.
The LBBE bounds continue to decrease with more servers, but some become inconsistent in so doing, which then actually increases the RMSE.

%
%
%


\miniskip\noindent\textbf{Asymmetry diversity.}  \ For fixed $\Rm$ values, the degree of bound tightening is critically dependent on the asymmetry diversity on the paths to the servers a client uses. 
The diversity we have used so far comes from the model established in Section~\ref{ssec:model}. 
We now explore the impact of stronger diversity to gain a more complete understanding of the benefits in a more general setting where the model may not hold, and also so we can explore up to very high levels of (relative) asymmetry {\dv so as to gain a complete picture.}
For other parameters the default values are used:  $M=16$, the 20 closest servers, 
and $\Rm$ values generated as usual. 
We work in the special case of zero congestion in order to examine of role of asymmetry diversity in isolation.

We use the following parametric family of symmetric distributions for $\Am$, which continues to obey the structure $\Am = \Rm\, T$. 
The (c,s) great circle SoL bounds result in $|T|$ taking values in $[0,T_L]$, where $T_L=1-2\dsol/\Rm\le 1$. 
As this is different for each (c,s) pair it is convenient to normalise to $[0,1]$ via $Z=|T|/T_L$.  We can now control \emph{achievable} asymmetry diversity by selecting $Z$ values uniformly over a range $I$ contained within $[0,1]$.
The resulting model for $T$ is then
 $T=\textit{sign}\cdot Z\cdot T_L$, where \textit{sign} is selected randomly from $\{-1,1\}$ with equal probability.   
 
We define five asymmetry levels according to the values of $I$ given in Table~\ref{tab:rmseAsymmetry}, in addition to the original {\dv empirical}  model-based values.
Regarding the latter,
 since in our simulation $\Rm$ is generated via 
 $\Rm= F\cdot2\dsol$ with $F\in[1.2,1.8]$,  we have $T_L=1-1/F\in[1/6,4/9]=[0.16^o,0.44^o]$, with an average value of $0.305$.

The RMSE results in the table confirm the expectation that error can be driven down provided sufficient asymmetry diversity is present.  However, to drop below a benchmark of 10\mus, we need to proceed to the asymmetry Level 5, 
a highly unrealistic level which would not be encountered in operational networks.
Similarly zero asymmetry ($T=Z=0$), which also implies zero error, is, as we know, not realistic.
\tinyskip

\begin{table}[h!]
\centering
    \scalebox{0.9}{
    \begin{tabular}{ c|c|c|c|c|c|c }
     &\!Model\!& Level 1\!& Level 2\!& Level 3\!& Level 4\!& Level 5\!\\ \hline
        I &\bb N/A \bb  &\bb$[0.6,0.7]$\bb &\bb[0.7,0.8]\bb &\bb$[0.8,0.9]$\bb  &\bb $[0.9,1]$\bb  &\bb$[0.99, 1]$\bb\\ \hline 
    \bb SBBE & 190 & 163 & 119 & 71 & 20 & 2\\
    \bb LBBE & 311 & 178 & 128 & 78 & 25 & 3\\
    
    \end{tabular} }
\vspace{4mm}
	\caption{RMSE (\mus) as a function of asymmetry level.}
	\label{tab:rmseAsymmetry}
\end{table}

\miniskip\noindent\textbf{Client geolocation errors.} \  For both methods clients and servers make available their geolocation as an input. While stratum-1 servers can be expected to know their location accurately, clients will in general have some error.  We expect however results to be insensitive to this, as an error of 1km corresponds, at $2c/3$, to only $5.0$\mus in the worst case.  Repeating our experiments with geolocation errors added to all clients, we found that {\dv the changes in average offset errors} were bounded by 1\mus for location errors below 1km.
Hence clients could simply obtain their geolocation from their street address.

\section{Generalizing SBBE to K-SBBE}
\label{sec:KSBBE}
The SBBE method can be readily generalized to incorporate partial routing information. The basic idea is simple:  if it were known that the $c\rightarrow s$ path passed through a localizable node $a$, then $c\rightarrow a\rightarrow s$ could be approximated by two great circles, each with a SoL bound, resulting in a larger, tighter lower bound on the OWD.

We establish our route ground truth as follows.
Consider a $(c,s)$ pair, set up as in 
Section~\ref{ssec:simulation}, with global path parameters $(\Rm,\Am;\dfm,\dbm)$. 
For the forward $c\rightarrow s$ path we can define a consistent underlying route with $N$ intermediate nodes as follows. For simplicity we express `distances' on the Earth's surface as latencies by scaling by the SoL in fibre:\\
\miniskip
\noindent \textbf{(i)} Place N nodes equidistantly on the $c\rightarrow s$ great circle, \\
\noindent \textbf{(ii)} select a random relative displacement $d_i$ for each node $i$ such that $\sum_i d_i =1$, and a random orientation $o_i\in\{N,S\}$,\\
\noindent \textbf{(iii)} given a perturbation amplitude $P$, perturb node $i$ by a distance $\dv P\cdot\!\dsol\!\cdot\!d_i$ in the direction $o_i$ at right angles to the great circle, and record the resulting perturbed {\dv delay} $D(P)$ as the sum of the $N+1$ per-hop great circle distances,\\
\noindent \textbf{(iv)} finalize the route by finding (numerically), the unique amplitude $P=P^*$ such that $D(P^*)= \dfm$.

\smallskip
The above procedure is repeated for the $s\rightarrow c$ direction to determine a reverse path with OWD matching $\dbm$.
For simplicity, we use the same value of $N$, as well as the same $\{d_i\}$ and $\{o_i\}$, ensuring that {\dv the paths in each direction} roughly follow each other.

\begin{figure}[b!]
\vspace{-8mm}
 \centering
   \hspace*{-05mm} 
   \includegraphics[trim={25mm 0 0 0mm},clip,width=80mm]{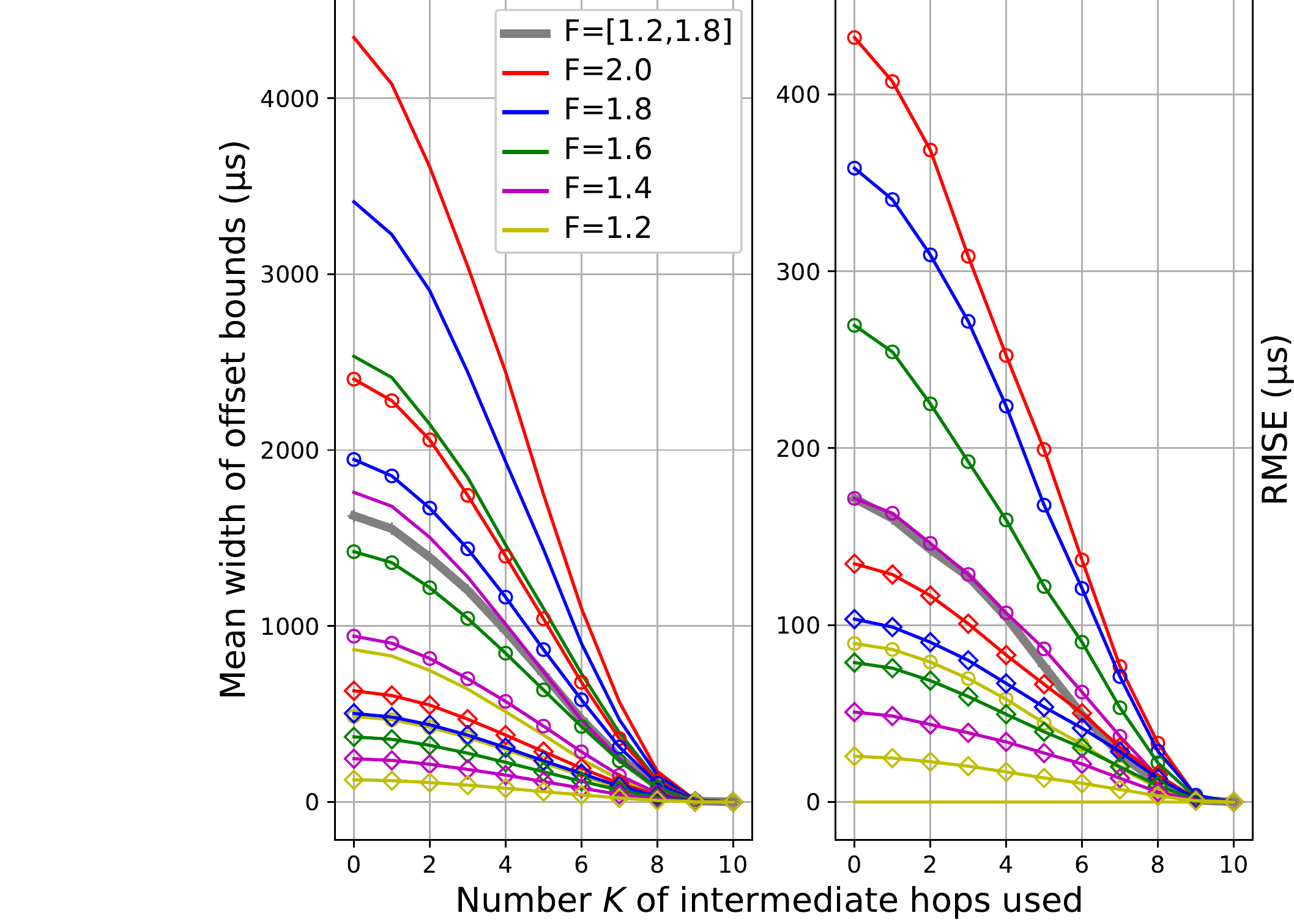}
  \includegraphics[trim={37mm 50mm 50mm 19mm},clip,width=61mm]{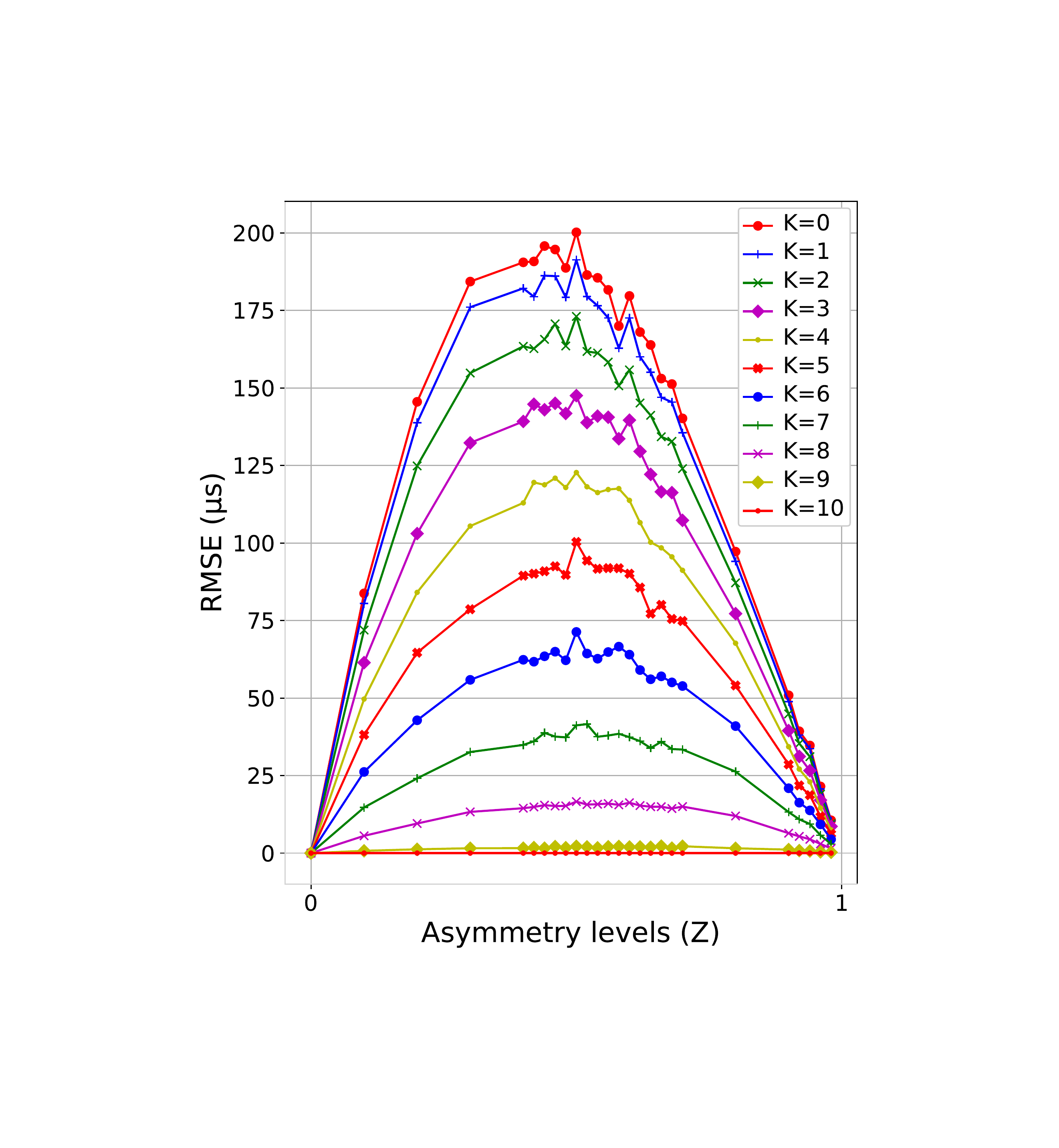}
   	\bigskip
	\caption{\dv Offset bound width (left plot) and RMSE (middle) as a function of $K$, for different route parameters (legend applies to both plots).
 	Asymmetry levels $Z=\{0,0.6,0.9\}$ are distinguished via plotting symbol \{none, circle, diamond\}.
 	Right plot: $Z$ dependence of error with $F=1.4$. The dependence is roughly unimodal and symmetric for each $K$.}
 	\label{fig:KSBBEbounds}
\end{figure}

We will assume the client is given the locations and order of $K\ge0$ of the $N$ intermediate nodes on each path, and that these are the `same' in each direction (same node indices).
The client can then form SoL lower bounds {\dv  $\dfb$ and $\dbb$ obeying} $\dsol\le\dfb\le\dfm$, $\dsol\le\dbb\le\dbm$ based on summing, in each direction separately, the $K+1$ great circle `hop' distances between the known nodes.
When $K=0$ the above reduces to the SBBE case where $\dfb=\dbb=\dsol$, however $\dfb\ne\dbb$ when $K>0$.
If $K=N$, then the round-trip route is fully known, and so the asymmetry bound width is zero since $\dfb=\dfm$ and $\dbb=\dbm$.


In the results to follow, we assume that each client connects to its 20 closest servers, using the same value of $K=10$ for each. 
As before, results are averaged over all clients, and zero congestion assumed.
We explore the $(\Rm,\Am)$ space by selecting
$\Rm = F\cdot2\dsol$ using inflation factors  $F=\{1.2,1.4,1.6,1.8,2.0\}$, and for each of these, asymmetries corresponding to $Z=\{0,0.6,0.9\}$ using the SoL-bound respecting parameterization of the previous section.
In addition, the result using the parameters
from Section~\ref{ssec:sensitivity} ($F\in[1.2,1.8]$, model based $\Am$), is included for comparison.
Note that large $Z$ does not necessarily mean high asymmetry in an absolute sense, but rather extreme asymmetry relative to what is \textit{possible}, given the value of $\Rm$ and the SoL bounds.

\tinyskip
Figure~\ref{fig:KSBBEbounds} (left plot) shows how the bound width drops with $K$, each curve being for a different $(\Rm,\Am)$ parameterized by $(F,Z)$.
Each curve is monotonic decreasing in $K$ because of the following: for each $(c,s)$ pair, when $K=1$ the node is chosen randomly. When increasing to $K=2$, the first node is retained, and a second randomly chosen from the remainder, and so on. 
The monotonic result reflects the expected strict tightening of bounds as routing knowledge increases.
We also see the expected monotonic behaviour w.r.t.~the other parameters:  
for $F$ fixed (same color) the curves are nested w.r.t.~$Z$, and
for $Z$ fixed (same symbol) the curves are nested w.r.t.~$F$, with tighter bounds in the direction of greater asymmetry and shorter routes respectively.
We observe that bounds are roughly halved when half the route is known.

{\dv The middle plot in} Figure~\ref{fig:KSBBEbounds} gives the corresponding results for average offset error.
The error dependence w.r.t.~$K$ and $F$ is analogous to that of the bound width, as expected. 
The dependence w.r.t.~asymmetry however is quite different: errors are zero, for all $K$ and $F$, in the zero asymmetry case ($Z=0$), where our estimator, defined as the centre of the bound, is exactly correct. 
On the other hand full asymmetry ($Z=1$, not shown), would also result in zero error as a byproduct of zero bound width. 
There is a trade off as higher asymmetry tightens bounds, but also pushes the true values toward the bound edges, initially increasing error. 

{\dv To explore this non-linear behaviour in $Z$ more fully, the right hand plot provides, for $F=1.4$, plots of error as a function of $Z$, on a much finer grid. 
For each value of $K$ we find a curve with a similar overall unimodal shape, with error vanishing at $Z=\{0,1\}$ as expected, and a peak at  $Z=Z^*(K)$ which varies only slightly with $K$, from $\approx 0.5$ for $K=0$, up to $\approx 0.6$ for $K=N$. The shapes of these curves for other $F$ values are very similiar. 
The irregularities observed in each curve, which are more prominent around the peak and which reduce as $K$ increases, were unexpected. We suspect they reflect the spatial heterogeneity of the (client, server) pairs averaged here.}


Finally, we note that to achieve errors in the tens of \mus range with {\dv this worst case achievable relative asymmetry ($Z^*=0.6$), low values of $F$ are required.  Based on the location of the empirical curve in the middle plot in Figure~\ref{fig:KSBBEbounds}, it seems that the average value in the network is not far from this unfortunate value.}
Errors reduce to zero when $K=N$, but just as for the bound width, significant reductions (on average) are not achieved
by only knowing one or two nodes.

\section{Deployment}
\label{sec:deployment} Broad deployment of our methods  would require client time synchronization software updates, and server updates in the case of LBBE landmark servers.    We envision the client-side implementation as an extension to NTP rather than a replacement since we only focus on one (critical) aspect of client synchronization.    Similarly, we envision a server-side implementation  as an extension of current server software.  Since there are a relatively small number of stratum 1 NTP servers in the Internet, incremental deployment should be straight-forward.  We plan to develop reference implementations for both clients and servers that can be tested in a live deployment in future work.

Client time synchronization software that implements our method would either need to include a preconfigured set of reference servers (this is common in NTP) or a service similar to  NTP Pool~\cite{ntppoolproject} that redirects client request to available reference servers, or the landmark network.  While it would add some complexity to do so, we argue that the latter is preferable since the redirection service could be designed to match clients to servers that are nearby (similar in spirit to redirection used in CDNs).

While our method does not require any change to time server infrastructure ({\em i.e.,} the stratum hierarchy), it does require additional configuration to establish measurements between landmark servers, and access to routing databases and IP path tracing to determine and track the nodes needed by K-SBBE.
{\dv Although it is beyond the scope of this paper to describe in detail how it would be done, there is no impediment to reliably identifying (and geo-locating) a number of the more prominent and stable landmarks along a route ({\em e.g.,} in IXPs or co-location centers).}
We envision configurations to be relatively stable.

Our method only adds a small amount of traffic to the existing timing infrastructure.  NTP clients can and already do probe multiple servers on an on-going basis.  Thus, the only additional traffic would be the inter-landmark measurements, which we argue is trivial.

\section{Summary: Toward 1\mus synchronization}
\label{sec:summary} Our study addresses the problem of client offset estimation in Internet time synchronization, focusing on its least tractable component: path asymmetry.  Assuming reasonable quality client clock algorithms, reducing asymmetry induced error is the \emph{essential requirement} to bring Internet client clock synchronization from milliseconds (the current standard) to microseconds, which is our goal.


We provide a theoretical framework that describes the direct relationship between offset error and the underlying path characteristics of the minimal RTT, and path asymmetry.  We quantify the extent of asymmetry in the Internet {\dv by exploiting a unique} empirical study based on high-precision OWD measurements. {\dv Although it has limitations, the result is the most detailed and accurate view of Internet asymmetry we are aware of.} These results are basis for an empirical model that we use to explore the likely impact of {\em asymmetry jitter} and {\em bound tightening}. {\dv We show} how the former can inflate offset and how latter can decrease errors considerably, but not enough to reach our goal. {\dv These findings are generic. They are informed by, but not fundamentally dependent upon, the details of the empirical model.}  Since this barrier is intrinsic, we turned to two models of extrinsic information to proceed further.

In SBBE, we exploit physical SoL limits to provide rigorous bounds that can tighten asymmetry intervals from individual client server pairs, magnifying the benefits of bound tightening. 
In LBBE, landmark servers provide clients with approximate delay bound information derived from reliable measurement of the inter-landmark network, plus SoL bounds.
For each method bound reductions were significant, but still insufficient.  We show that server density, which controls the $\Rm$ of the closest server to clients, is the key controlling factor. 

To proceed further without the implied cost of increasing server density, we extended SBBE to K-SBBE, where a client has access to the location of $K$ nodes along the route (in each direction) to servers.  
This offers an in-principle pathway to zero offset error, for sufficiently large $K$. 
{\dv We also discovered and explained the non-linear relationship of asymmetry induced error as a function of asymmetry, and characterized the worst case.}  
However we found that low perturbations from SoL, and/or route information well beyond just one or two nodes, are needed to enter the sub-10\mus regime for average clients, who cannot be assumed to be only 1ms or less away from stratum-1 servers, or to enjoy either extreme, or zero, path asymmetry. 

To summarise, to move toward a practical 10\mus benchmark requires, in addition to bound tightening techniques and SoL bounds, more detailed knowledge of (calibrated) network routing by clients or landmarks, together with close servers, with server diversity being, finally, less important. To move to 1\mus would require, in addition, calibration within client operating systems themselves, and remains very challenging in the Internet context, though not impossible.

\bibliographystyle{ACM-Reference-Format}
\bibliography{paper}


\begin{thebibliography}{25}


\ifx \showCODEN    \undefined \def \showCODEN     #1{\unskip}     \fi
\ifx \showDOI      \undefined \def \showDOI       #1{#1}\fi
\ifx \showISBNx    \undefined \def \showISBNx     #1{\unskip}     \fi
\ifx \showISBNxiii \undefined \def \showISBNxiii  #1{\unskip}     \fi
\ifx \showISSN     \undefined \def \showISSN      #1{\unskip}     \fi
\ifx \showLCCN     \undefined \def \showLCCN      #1{\unskip}     \fi
\ifx \shownote     \undefined \def \shownote      #1{#1}          \fi
\ifx \showarticletitle \undefined \def \showarticletitle #1{#1}   \fi
\ifx \showURL      \undefined \def \showURL       {\relax}        \fi
\providecommand\bibfield[2]{#2}
\providecommand\bibinfo[2]{#2}
\providecommand\natexlab[1]{#1}
\providecommand\showeprint[2][]{arXiv:#2}

\bibitem[\protect\citeauthoryear{Cao and Veitch}{Cao and Veitch}{2018}]%
        {sync_Leap2016}
\bibfield{author}{\bibinfo{person}{Y. Cao} {and} \bibinfo{person}{D. Veitch}.}
  \bibinfo{year}{2018}\natexlab{}.
\newblock \showarticletitle{{Network Timing, Weathering the 2016 Leap Second}}.
  In \bibinfo{booktitle}{\emph{{Proceedings of IEEE INFOCOM}}}.
  \bibinfo{address}{Honolulu, USA}.
\newblock


\bibitem[\protect\citeauthoryear{Cao and Veitch}{Cao and Veitch}{2019a}]%
        {sync_TimeServer_Dataset2016-17}
\bibfield{author}{\bibinfo{person}{Y. Cao} {and} \bibinfo{person}{D. Veitch}.}
  \bibinfo{year}{2019}\natexlab{a}.
\newblock \bibinfo{title}{{TimeServer Dataset 2016-2017}}.
\newblock
  \bibinfo{howpublished}{\url{https://data.research.uts.edu.au/public/DVTSD/}}.
\newblock
\urldef\tempurl%
\url{https://doi.org/10.26195/5bf790bd1b6a0}
\showDOI{\tempurl}


\bibitem[\protect\citeauthoryear{Cao and Veitch}{Cao and Veitch}{2019b}]%
        {sync_Best50}
\bibfield{author}{\bibinfo{person}{Yi Cao} {and} \bibinfo{person}{Darryl
  Veitch}.} \bibinfo{year}{2019}\natexlab{b}.
\newblock \showarticletitle{{Where on Earth are the Best-50 Time Servers?}}. In
  \bibinfo{booktitle}{\emph{Proceedings of the Passive and Active Measurement
  Conference}}. \bibinfo{address}{Puerto Varas, Chile}.
\newblock
\urldef\tempurl%
\url{http://www.crin.eng.uts.edu.au/~darryl/Publications/Best50_PAM2019_camera.pdf}
\showURL{%
\tempurl}


\bibitem[\protect\citeauthoryear{{D.L. Mills}}{{D.L. Mills}}{2012}]%
        {millsntp}
\bibfield{author}{\bibinfo{person}{{D.L. Mills}}.}
  \bibinfo{year}{2012}\natexlab{}.
\newblock \bibinfo{title}{{Computer Network Time Synchronization}}.
\newblock
  \bibinfo{howpublished}{\url{https://www.eecis.udel.edu/~mills/exec.html}}.
\newblock


\bibitem[\protect\citeauthoryear{Fraleigh, Moon, Lyles, Cotton, Khan, Moll,
  Rockell, Seely, and Diot}{Fraleigh et~al\mbox{.}}{2003}]%
        {Fraleigh2003}
\bibfield{author}{\bibinfo{person}{C. Fraleigh}, \bibinfo{person}{S. Moon},
  \bibinfo{person}{B. Lyles}, \bibinfo{person}{C. Cotton}, \bibinfo{person}{M.
  Khan}, \bibinfo{person}{D. Moll}, \bibinfo{person}{R. Rockell},
  \bibinfo{person}{T. Seely}, {and} \bibinfo{person}{C. Diot}.}
  \bibinfo{year}{2003}\natexlab{}.
\newblock \showarticletitle{{Packet-level Traffic Measurements from the Sprint
  IP Backbone}}.
\newblock \bibinfo{journal}{\emph{{IEEE Network}}} \bibinfo{volume}{17},
  \bibinfo{number}{6} (\bibinfo{date}{November} \bibinfo{year}{2003}).
\newblock


\bibitem[\protect\citeauthoryear{{Freris}, {Graham}, and {Kumar}}{{Freris}
  et~al\mbox{.}}{2011}]%
        {Kumar_TAC2010}
\bibfield{author}{\bibinfo{person}{N.~M. {Freris}}, \bibinfo{person}{S.~R.
  {Graham}}, {and} \bibinfo{person}{P.~R. {Kumar}}.}
  \bibinfo{year}{2011}\natexlab{}.
\newblock \showarticletitle{Fundamental Limits on Synchronizing Clocks Over
  Networks}.
\newblock \bibinfo{journal}{\emph{IEEE Trans. Automat. Control}}
  \bibinfo{volume}{56}, \bibinfo{number}{6} (\bibinfo{year}{2011}),
  \bibinfo{pages}{1352--1364}.
\newblock


\bibitem[\protect\citeauthoryear{Geng, Liu, Yin, Naik, Prabhakar, Rosenblum,
  and Vahdat}{Geng et~al\mbox{.}}{2018}]%
        {huygens}
\bibfield{author}{\bibinfo{person}{Y. Geng}, \bibinfo{person}{S. Liu},
  \bibinfo{person}{Z. Yin}, \bibinfo{person}{A. Naik}, \bibinfo{person}{B.
  Prabhakar}, \bibinfo{person}{M. Rosenblum}, {and} \bibinfo{person}{A.
  Vahdat}.} \bibinfo{year}{2018}\natexlab{}.
\newblock \showarticletitle{{Exploiting a Natural Network Effect for Scalable,
  Fine-grained Clock Synchronization}}. In
  \bibinfo{booktitle}{\emph{{Proceedings of the USENIX NSDI Symposium}}}.
  \bibinfo{address}{Renton, WA}.
\newblock


\bibitem[\protect\citeauthoryear{Hansen}{Hansen}{2019}]%
        {ntppoolproject}
\bibfield{author}{\bibinfo{person}{B. Hansen}.}
  \bibinfo{year}{2019}\natexlab{}.
\newblock \bibinfo{title}{{NTP Pool Project}}.
\newblock \bibinfo{howpublished}{\url{http://www.pool.ntp.org/en/}}.
\newblock


\bibitem[\protect\citeauthoryear{He, Faloutsos, Krishnamurthy, and Huffaker}{He
  et~al\mbox{.}}{2005}]%
        {He05}
\bibfield{author}{\bibinfo{person}{Y. He}, \bibinfo{person}{M. Faloutsos},
  \bibinfo{person}{S. Krishnamurthy}, {and} \bibinfo{person}{B. Huffaker}.}
  \bibinfo{year}{2005}\natexlab{}.
\newblock \showarticletitle{{On Routing Asymmetry in the Internet}}. In
  \bibinfo{booktitle}{\emph{{Proceedings of the IEEE Global Telecommunications
  Conference (Globecom)}}}. \bibinfo{address}{St. Louis, MO}.
\newblock


\bibitem[\protect\citeauthoryear{Hong, Lin, and Caesar.}{Hong
  et~al\mbox{.}}{2011}]%
        {Hong11}
\bibfield{author}{\bibinfo{person}{C. Hong}, \bibinfo{person}{C. Lin}, {and}
  \bibinfo{person}{M. Caesar.}} \bibinfo{year}{2011}\natexlab{}.
\newblock \showarticletitle{{Clockscalpel: Understanding Root Causes of
  Internet Clock Synchronization Inaccuracy}}. In
  \bibinfo{booktitle}{\emph{{Proceedings of the Passive and Active Network
  Measurement Conference}}}. \bibinfo{address}{Cleveland, OH}.
\newblock


\bibitem[\protect\citeauthoryear{John, Dusi, and Claffy}{John
  et~al\mbox{.}}{2010}]%
        {John2010}
\bibfield{author}{\bibinfo{person}{W. John}, \bibinfo{person}{M. Dusi}, {and}
  \bibinfo{person}{K.C. Claffy}.} \bibinfo{year}{2010}\natexlab{}.
\newblock \showarticletitle{{Estimating Routing Symmetry on Single Links by
  Passive Flow Measurements}}. In \bibinfo{booktitle}{\emph{{Proceedings of the
  6th IEEE International Wireless Communications and Mobile Computing
  Conference}}}. \bibinfo{address}{Caen, France}.
\newblock


\bibitem[\protect\citeauthoryear{Mani, Durairajan, Barford, and Sommers}{Mani
  et~al\mbox{.}}{2018}]%
        {SpotIOT18}
\bibfield{author}{\bibinfo{person}{S.K. Mani}, \bibinfo{person}{R. Durairajan},
  \bibinfo{person}{P. Barford}, {and} \bibinfo{person}{J. Sommers}.}
  \bibinfo{year}{2018}\natexlab{}.
\newblock \showarticletitle{{An Architecture for IoT Clock Synchronization}}.
  In \bibinfo{booktitle}{\emph{Proceedings of the International Conference on
  the Internet of Things}}. \bibinfo{address}{Santa Barbara, CA}.
\newblock


\bibitem[\protect\citeauthoryear{Marzullo and Owicki}{Marzullo and
  Owicki}{1983}]%
        {Marzullo83}
\bibfield{author}{\bibinfo{person}{K. Marzullo} {and} \bibinfo{person}{S.
  Owicki}.} \bibinfo{year}{1983}\natexlab{}.
\newblock \showarticletitle{{Maintaining the Time in a Distributed System}}. In
  \bibinfo{booktitle}{\emph{Proceedings of the Second Annual ACM Symposium on
  Principles of Distributed Computing}} (Montreal, Quebec, Canada).
  \bibinfo{pages}{295–305}.
\newblock
\urldef\tempurl%
\url{https://doi.org/10.1145/800221.806730}
\showURL{%
\tempurl}


\bibitem[\protect\citeauthoryear{Mills}{Mills}{1981}]%
        {mills81dcnet}
\bibfield{author}{\bibinfo{person}{D.L. Mills}.}
  \bibinfo{year}{1981}\natexlab{}.
\newblock \bibinfo{title}{{DCNET Internet Clock Service}}.
\newblock \bibinfo{howpublished}{\url{https://tools.ietf.org/html/rfc778}}.
\newblock


\bibitem[\protect\citeauthoryear{Mills}{Mills}{1985a}]%
        {mills85alg}
\bibfield{author}{\bibinfo{person}{D.L. Mills}.}
  \bibinfo{year}{1985}\natexlab{a}.
\newblock \bibinfo{title}{{Algorithms for Synchronizing Network Clocks}}.
\newblock \bibinfo{howpublished}{\url{https://tools.ietf.org/html/rfc956}}.
\newblock


\bibitem[\protect\citeauthoryear{Mills}{Mills}{1985b}]%
        {rfc958}
\bibfield{author}{\bibinfo{person}{D.L. Mills}.}
  \bibinfo{year}{1985}\natexlab{b}.
\newblock \bibinfo{title}{{Network Time Protocol (NTP)}}.
\newblock \bibinfo{howpublished}{\url{https://tools.ietf.org/html/rfc958}}.
\newblock


\bibitem[\protect\citeauthoryear{Mills, Martin, Burbank, and Kasch}{Mills
  et~al\mbox{.}}{2010}]%
        {rfc5905}
\bibfield{author}{\bibinfo{person}{D. Mills}, \bibinfo{person}{J. Martin},
  \bibinfo{person}{J. Burbank}, {and} \bibinfo{person}{W. Kasch}.}
  \bibinfo{year}{2010}\natexlab{}.
\newblock \bibinfo{title}{{Network Time Protocol Version 4: Protocol and
  Algorithms Specification}}.
\newblock \bibinfo{howpublished}{\url{https://tools.ietf.org/html/rfc5905}}.
\newblock


\bibitem[\protect\citeauthoryear{Pathak, Pucha, Zhang, C.Hu, and Mao}{Pathak
  et~al\mbox{.}}{2008}]%
        {Pathak2008}
\bibfield{author}{\bibinfo{person}{A. Pathak}, \bibinfo{person}{H. Pucha},
  \bibinfo{person}{Y. Zhang}, \bibinfo{person}{Y. C.Hu}, {and}
  \bibinfo{person}{Z.~M. Mao}.} \bibinfo{year}{2008}\natexlab{}.
\newblock \showarticletitle{{A Measurement Study of Internet Delay Asymmetry}}.
  In \bibinfo{booktitle}{\emph{{Proceedings of the Passive and Active Network
  Measurement Conference}}}. \bibinfo{address}{Cleveland, OH}.
\newblock


\bibitem[\protect\citeauthoryear{Paxson}{Paxson}{1997}]%
        {Vern97}
\bibfield{author}{\bibinfo{person}{V. Paxson}.}
  \bibinfo{year}{1997}\natexlab{}.
\newblock \showarticletitle{{End-to-end Routing Behavior in the Internet}}.
\newblock \bibinfo{journal}{\emph{IEEE/ACM Transactions on Networking}}
  \bibinfo{volume}{5}, \bibinfo{number}{5} (\bibinfo{date}{October}
  \bibinfo{year}{1997}).
\newblock


\bibitem[\protect\citeauthoryear{Schwartz, Shavitt, and Weinsberg}{Schwartz
  et~al\mbox{.}}{2010}]%
        {Schwartz10}
\bibfield{author}{\bibinfo{person}{Y. Schwartz}, \bibinfo{person}{Y. Shavitt},
  {and} \bibinfo{person}{U. Weinsberg}.} \bibinfo{year}{2010}\natexlab{}.
\newblock \showarticletitle{{On the Diversity, Stability and Symmetry of
  End-to-End Internet Routes}}. In \bibinfo{booktitle}{\emph{{Proceedings of
  the IEEE INFOCOM Conference}}}. \bibinfo{address}{San Diego, CA}.
\newblock


\bibitem[\protect\citeauthoryear{Veitch, Ridoux, and Korada}{Veitch
  et~al\mbox{.}}{2009}]%
        {sync_ToN}
\bibfield{author}{\bibinfo{person}{D. Veitch}, \bibinfo{person}{J. Ridoux},
  {and} \bibinfo{person}{S. Korada}.} \bibinfo{year}{2009}\natexlab{}.
\newblock \showarticletitle{{Robust Synchronization of Absolute and Difference
  Clocks over Networks}}.
\newblock \bibinfo{journal}{\emph{IEEE/ACM Transactions on Networking}}
  \bibinfo{volume}{17}, \bibinfo{number}{2} (\bibinfo{date}{April}
  \bibinfo{year}{2009}).
\newblock


\bibitem[\protect\citeauthoryear{Vijayalayan and Veitch}{Vijayalayan and
  Veitch}{2016}]%
        {sync_SHM}
\bibfield{author}{\bibinfo{person}{K. Vijayalayan} {and} \bibinfo{person}{D.
  Veitch}.} \bibinfo{year}{2016}\natexlab{}.
\newblock \showarticletitle{{Rot at the roots? Examining Public Timing
  Infrastructure}}. In \bibinfo{booktitle}{\emph{{Proceedings of the IEEE
  INFOCOM Conference}}}. \bibinfo{address}{San Francisco, CA}.
\newblock


\bibitem[\protect\citeauthoryear{Wassermann, Casas, Donnet, Leduc, and
  Mellia}{Wassermann et~al\mbox{.}}{2016}]%
        {Wassermann2016}
\bibfield{author}{\bibinfo{person}{S. Wassermann}, \bibinfo{person}{P. Casas},
  \bibinfo{person}{B. Donnet}, \bibinfo{person}{G. Leduc}, {and}
  \bibinfo{person}{M. Mellia}.} \bibinfo{year}{2016}\natexlab{}.
\newblock \showarticletitle{{On the Analysis of Internet Paths with DisNETPerf,
  a Distributed Paths Performance Analyzer}}. In
  \bibinfo{booktitle}{\emph{{Proceedings of the 41st IEEE Conference on Local
  Computer Networks Workshops}}}. \bibinfo{address}{Dubai, UAE}.
\newblock


\bibitem[\protect\citeauthoryear{Wong, Stoyanov, and Sirer}{Wong
  et~al\mbox{.}}{2007}]%
        {Wong07}
\bibfield{author}{\bibinfo{person}{B. Wong}, \bibinfo{person}{I. Stoyanov},
  {and} \bibinfo{person}{E. Sirer}.} \bibinfo{year}{2007}\natexlab{}.
\newblock \showarticletitle{{Octant: A Comprehensive Framework for the
  Geolocation of Internet Hosts}}. In \bibinfo{booktitle}{\emph{Proceedings of
  the {USENIX} Symposium on Networked Systems Design and Implementation}}.
  \bibinfo{address}{Cambridge, MA}.
\newblock


\bibitem[\protect\citeauthoryear{Zeitoun, Chuah, Bhattacharyya, and
  Diot}{Zeitoun et~al\mbox{.}}{2004}]%
        {Zeitoun2004}
\bibfield{author}{\bibinfo{person}{A. Zeitoun}, \bibinfo{person}{C. Chuah},
  \bibinfo{person}{S. Bhattacharyya}, {and} \bibinfo{person}{C. Diot}.}
  \bibinfo{year}{2004}\natexlab{}.
\newblock \showarticletitle{{An AS-level Study of Internet Path Delay
  Characteristics}}. In \bibinfo{booktitle}{\emph{{Proceedings of IEEE Global
  Telecommunications Conference}}}. \bibinfo{address}{Dallas, TX}.
\newblock


\end{thebibliography}

\newpage
\section{Appendix}
\label{sec:appendix} For interest, we provide some additional zoomed views of the scatterplot of measured relative asymmetry $T$ against $\Rm$ displayed in full in Figure~\ref{fig:scatter}.
Each point corresponding to a $(\Rm,\Am)$ measurement made across a single Clear Zone. Points from the same server share the same color.

\begin{figure*}[h!]  
    \begin{center}
    {\includegraphics[width=140mm]{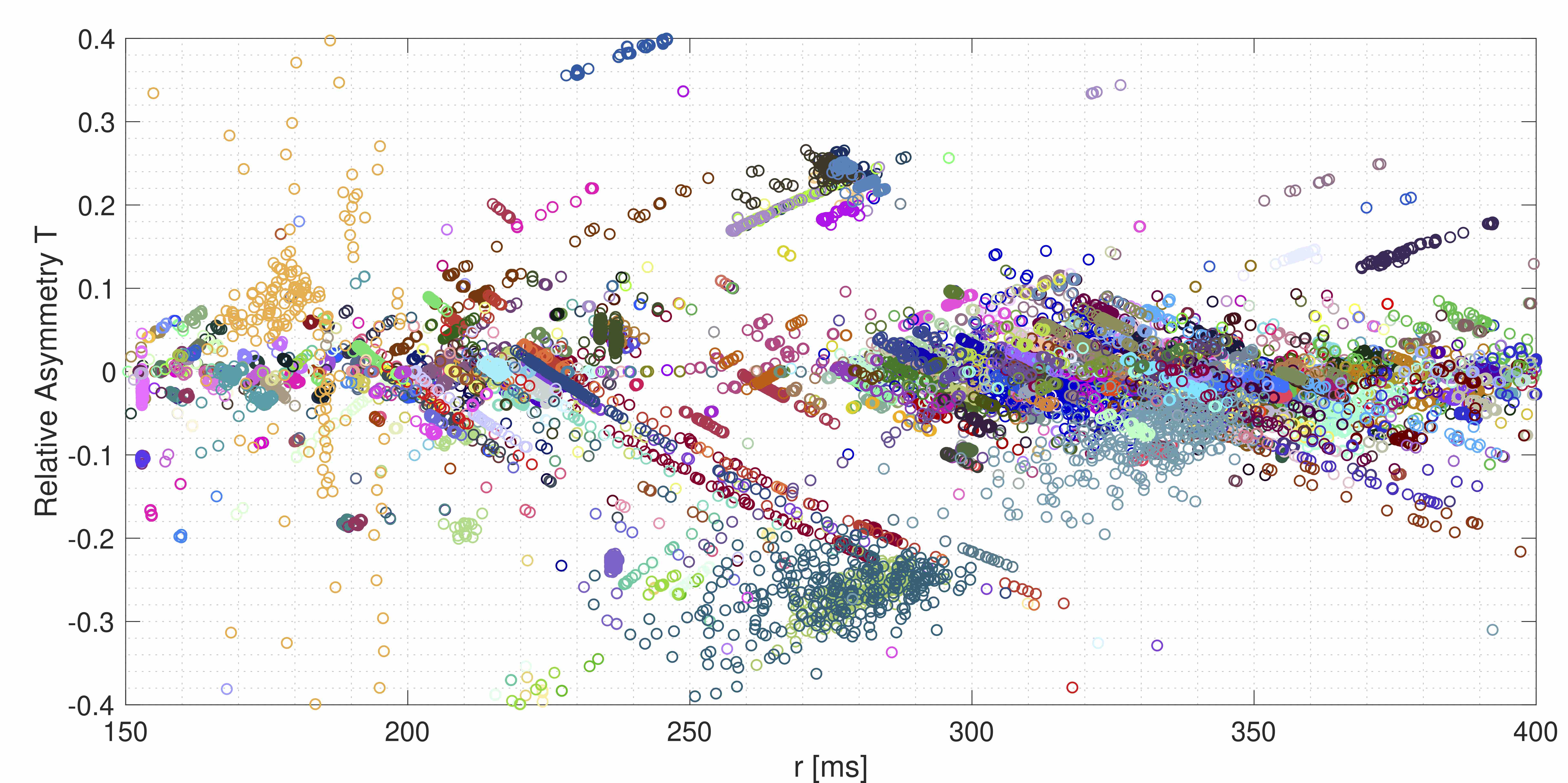}}
    \end{center}
    \caption{Central zoom showing the loose transition gap around 280\ms corresponding, roughly, to US versus European servers .
    }
    \label{fig:scatter_zoomcentral}
\end{figure*}

\begin{figure*}[h!]  
    \begin{center}
    {\includegraphics[width=140mm]{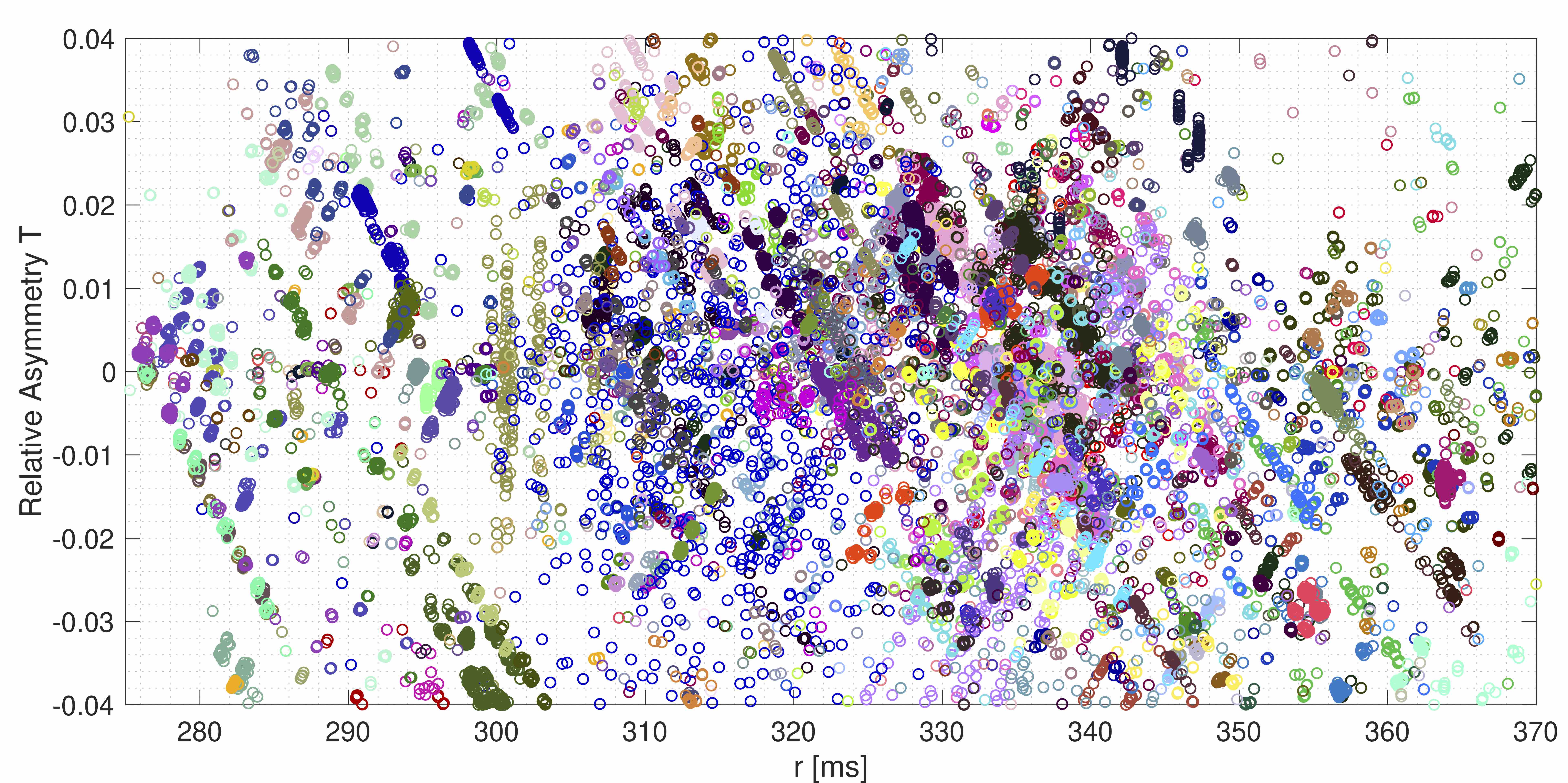}}
    \end{center}
    \caption{Closer zoom showing the central body of values, mainly from the US.
    }
    \label{fig:scatter_USheart}
\end{figure*}

\end{document}